\documentclass[twoside,twocolumn]{article}

\usepackage[utf8]{inputenc}
\usepackage[sc]{mathpazo} 
\usepackage[T1]{fontenc} 
\usepackage{lmodern} 
\usepackage{microtype} 

\usepackage[english]{babel} 

\usepackage{graphicx}
\usepackage{color}
\usepackage[top=2cm,bottom=2cm,left=1.5cm,right=1.5cm,columnsep=20pt]{geometry} 
    {\large\scshape Computational Details%
    \par\medskip\normalfont\normalsize}%
    {}%
    {\large\scshape Acknowledgement%
    \par\medskip\normalfont\normalsize}%
    {}%
    {\large\scshape Supporting Information%
    \par\medskip\normalfont\normalsize}%
    {}%
\usepackage{multirow} 
\usepackage[normal, small,labelfont=bf,up,textfont=it,up]{caption} 
\usepackage{booktabs} 

\usepackage{lettrine} 

\usepackage{enumitem} 
\setlist[itemize]{noitemsep} 

\usepackage{abstract} 

\usepackage{titlesec} 
\renewcommand\thesection{\Roman{section}} 
\renewcommand\thesubsection{\roman{subsection}} 
\titleformat{\section}[block]{\large\scshape\centering}{\thesection.}{1em}{} 
\titleformat{\subsection}[block]{\large}{\thesubsection.}{1em}{} 

\usepackage{fancyhdr} 
\usepackage{titling} 

\usepackage{hyperref} 

\pretitle{\begin{center}\Large\bfseries} 
\posttitle{\end{center}} 

\def\bea{\begin{eqnarray}}
\def\eea{\end{eqnarray}}
\def\ben{\begin{equation}}
\def\een{\end{equation}}
\def\benu{\begin{enumerate}}
\def\enu{\end{enumerate}}

\def\bei{\begin{itemize}}
\def\eei{\end{itemize}}
\def\beit{\begin{itemize}}
\def\eit{\end{itemize}}
\def\benu{\begin{enumerate}}
\def\enu{\end{enumerate}}

\def\sss{\scriptscriptstyle\rm}

\def\1var{(\bx_1...\bx\N)}

\def\bx{{x}}

\def\N{_{\sss N}}

\def\sph_int{ {\int d^3 r}}

\usepackage[table]{xcolor}
\usepackage{amsmath}
\usepackage{graphicx}

\usepackage[]{quoting}
\usepackage{multirow} 
\usepackage[table]{xcolor}

\def\d{_{\sss D}}
\def\f{_{\sss F}}

\title{Density-corrected DFT explained:  Questions and answers}

\author{%
\textsc{Suhwan Song$^a$, Stefan Vuckovic$^{b,c}$, Eunji Sim$^{a,}$\thanks{esim@yonsei.ac.kr}, and Kieron Burke$^d$} \\ 
\normalsize $^a$Department of Chemistry, Yonsei University, 50 Yonsei-ro Seodaemun-gu, Seoul 03722, Korea \\ 
\normalsize $^b$Institute for Microelectronics and Microsystems (CNR-IMM), Via Monteroni,Campus Unisalento, 73100 Lecce, Italy \\
\normalsize $^c$Department of Chemistry\&Pharmaceutical Sciences and Amsterdam Institute of Molecular and Life Sciences (AIMMS), \\
\normalsize Faculty of Science, Vrije Universiteit, De Boelelaan 1083, 1081HV Amsterdam, The Netherlands\\
\normalsize $^d$Departments of Chemistry and of Physics, University of California, Irvine, CA 92697, USA \\  
}
\date{\today} 

\usepackage{xr}
\usepackage{float}

\renewcommand{\maketitlehookd}{%
\begin{abstract}
\noindent 
HF-DFT, the practice of evaluating approximate density functionals on Hartree-Fock densities, has long been used in testing density
functional approximations.
Density-corrected DFT (DC-DFT) is a general theoretical framework for identifying 
failures of density functional approximations by
separating errors in a functional from errors in its 
self-consistent (SC) density.
Most modern DFT calculations yield highly accurate densities, but important characteristic classes of calculation have large
density-driven errors,
including reaction barrier heights, electron affinities, radicals and anions in solution, dissociation of heterodimers, and even some torsional barriers.
Here, the HF density (if not spin-contaminated) usually yields more accurate and consistent energies than those of the 
SC density.
We use the term DC(HF)-DFT to indicate DC-DFT using HF densities only in such cases.
A recent comprehensive study 
(J. Chem. Theory Comput. 2021, 17, 1368$-$1379)
of HF-DFT led to many unfavorable conclusions.   
A re-analysis using DC-DFT shows 
that DC(HF)-DFT substantially improves
DFT results precisely when
SC densities are flawed.
\end{abstract}
}

\begin{document}

\maketitle
\sf
\small

\section{Introduction}

In any practical Kohn-Sham density functional theory (KS-DFT) calculation, of which there are tens of thousands
reported each year, the exchange-correlation (XC) is approximated, and there
are literally hundreds of approximations in common use today.\cite{MH17}
In every single case, one can consider the error in the calculation as arising
from two different sources:  the error in the self-consistent (SC) density
and the error in the final evaluation of the total energy.\cite{WNJK17}   
Density-corrected DFT (DC-DFT) is a formal framework 
for distinguishing these two types of error.\cite{KSB13, KSB14}

To be clear, in the vast majority of DFT calculations reported, the
SC density is an excellent approximation
to the exact density, so that the
error in the energy is almost entirely due to the functional approximation
to the energy.
However, in a significant subset of calculations where standard
approximations show surprisingly large errors, these errors can be
traced to the error in the SC density.\cite{KPSS15, KSSB18}
In such cases, use of a more
accurate density usually reduces the error significantly (typically
by a factor of 2 or more).  
This has been demonstrated for specific
classes of problems, such as electron affinities\cite{KSB11}, 
dissociation energy curve of diatomic molecules\cite{KPSS15,NSSB20},
radical ions in solution\cite{KSB14},
$\sigma$-hole interaction energies of halogen bonds\cite{KSSB18},
and spin gaps of transition metal complexes\cite{SKSB18}.

This observation would be useless if one always needed highly accurate 
densities to generate these improvements in energies, as such densities
would cost at least as much as generating highly accurate energies
via, e.g., CCSD(T).\cite{NSSB20}  However, in practice, when standard semilocal
density functionals fail due to density-driven errors, 
the Hartree-Fock (HF) density is usually sufficient to generate
energetic improvements that are comparable to those of the exact density.
In various calculations going back to the early 1990's and the rise
of DFT for computation in chemistry, there have been applications
of this approach, dubbed HF-DFT, mostly for convenience,
but no general understanding of when it should be
applied.\cite{CC90, GJPF92,GJPF92b, HMAA92, S92,JGP92,OB94}
Once codes were reworked to automatically perform self-consistent DFT
calculations, which has both formal and practical advantages, it became largely obsolete.

DC-DFT is the theory that explains when HF-DFT should
be applied and when it should work.
The significance of DC-DFT is that it helps the systematic development
of density functional approximation by providing a principle for determining the true error. 
The exact functional is free of density errors, since both its
energy and its functional derivative (which determines the density) are exact. 
For {\em any} approximation, the error in the energy has contributions from both the functional 
and the density.
Typical DFT error analysis is based on the total energy error and does not distinguish the
source of the error nor determine the accuracy of the functional. 
According to the principles of DC-DFT, researchers should use just functional errors 
(or use them together with density-driven errors) to formulate strategies and guidelines
for functional correction and development.  In the early days of DFT in quantum chemistry,
these principles were largely irrelevant, as functionals were rather crude and most
calculations had relatively negligible density-driven errors (hence the
use of HF-DFT when convenient).  With the rise in applications of DFT to
many different systems throughout chemistry\cite{J15,RH16,GHBE17}, 
the characterization of errors for specific properties or systems,
and the ever increasing accuracy of modern functionals\cite{MH17,MH16},
the importance of the principles of 
DC-DFT has grown enormously.

\begin{figure*}[htb]
\centering
\includegraphics[width=0.45\linewidth]{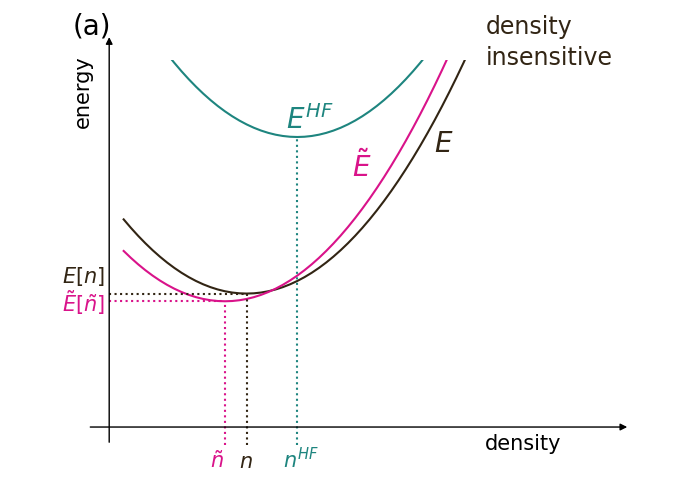}
\includegraphics[width=0.45\linewidth]{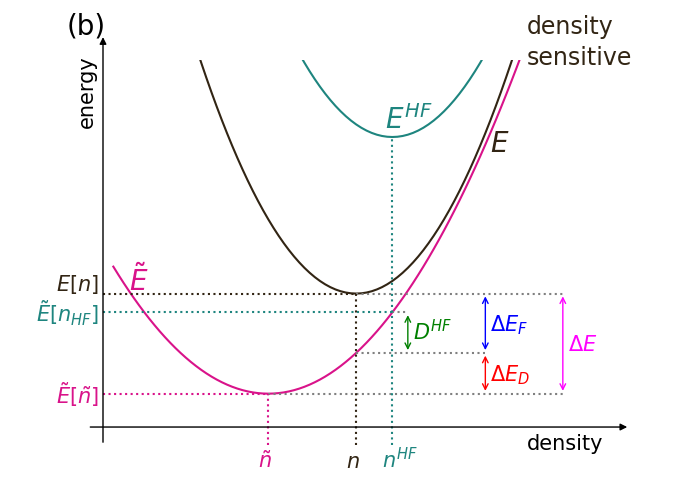}
\captionof{figure}[Cartoon of DC-DFT idea]{
Cartoon of the exact (black),   approximate (magenta,
denoted with tilde) and HF energy functionals (green). 
(a) is the usual case, where accuracy is not much
affected by which density is used;
(b) is a density-sensitive case,
where use of exact (or HF) density significantly reduces error.
$\tilde{E}[n^{HF}]$ is the approximate energy
on the HF density,
while $\tilde{E}[\tilde{n}]$ is its self-consistent ground-state energy.
}
\label{fgr:cartoon}
\end{figure*}

DC-DFT is a rigorous theory, separating errors using the exact density.
It identifies approximate DFT calculations as abnormal when the density-driven
error is signficant.  In practice, it is most often applied using the 
HF density in place of the exact one, and using a simple heuristic like
density sensitivity in place of abnormality to decide if the HF density
should be used in place of the self-consistent density.   We have called
that procedure DC(HF)-DFT, meaning density-corrected DFT, using the HF
density but only when it is appropriate.

The most immediate application of DC(HF)-DFT is to understand HF-DFT, by
producing a recipe stating when it should improve over SC-DFT.
Thus, DC(HF)-DFT includes two criteria for when
a HF density should be used instead of a SC density.
The most important is that the calculation under consideration should
be contaminated by a density-driven error.  Simple measures that
are suitable for many circumstances have evolved.\cite{SSB18}  The current
criteria is if the energy calculation shows
pronounced density sensitivity, i.e., the result changes appreciably
when evaluated on two distinct densities.   
Second, because the HF density 
is a proxy for the exact density,
we avoid its use when it is spin contaminated (note that spin contamination is typically less of a problem for SC-DFT).\cite{PRCS09}
These criteria are both common sense and understandable
in terms of the general theory of DC-DFT.

This paper explains the current state of the art of DC(HF)-DFT, and its
relation to HF-DFT (meaning always using HF densities instead of self-consistent ones).
In the background section,
we give both the general theory of DC-DFT and the history of HF-DFT.
The recent work of 
Jan Martin and co-workers in Ref.~\cite{SM21} 
was possibly the most comprehensive study
of HF-DFT to date (although, as we detailed below, 
it contained several conceptual confusions and other difficulties).
But it did not apply DC(HF)-DFT
methodology in its current form.
We illustrate with many examples from Ref.~\cite{SM21} data the
detailed performance of DC(HF)-DFT.

\section{Background}

\subsection{DC-DFT}

The core idea of DC-DFT is to separate the energetic error of any 
SC-DFT calculation into
a functional contribution and a density-driven error,  where the latter vanishes if
an approximate functional is evaluated on the exact density. Writing the total SC-DFT energy error as: $\Delta E = \tilde E [\tilde n] - E[n]$, where $ E$ and $ n$ are the exact 
energy functional and density, and $\tilde E$ and $\tilde n$ are their approximate counterparts, $\Delta E$ is decomposed as:\cite{KSB13,NSSB20,SSB18}:
\ben
\Delta E=\underbrace{\tilde E[\tilde n] - \tilde E[n]}_{\Delta E\d} 
+
 \underbrace{ \tilde E[n] - E[n]}_{\Delta E\f}, 
\label{eq:FD}
\een
where $\Delta E\f$ is the functional error and
$\Delta E\d$ is the 
density-driven error.  
Almost all present DFT calculations use the 
KS scheme\cite{KS65}, in which the
functional error is simply the error 
in the XC approximation, evaluated on
the exact density.  
Moreover, almost all such calculations 
are for energy differences, 
so one
applies Eq.~\ref{eq:FD} to each calculation separately, 
and subtracts.

Most standard XC approximations begin with a semi-local form, including a generalized gradient approximation (GGA).
For many molecules and properties of chemical interest, such as dissociations energies,
the magnitude of $\Delta E\d$ with such functionals is much smaller than that of $\Delta E\f$, and can be neglected.
On the other hand, in density-sensitive calculations, $\left| \Delta E\d \right| $ represents
a significant fraction of the total error (see the cartoons in Fig.~\ref{fgr:cartoon} comparing density-sensitive and insensitive calculations).
The density-sensitivity of a DFT calculation depends on the system under study, the property being extracted, and the
approximate functional.

Generically (but not always),  removal of the density-driven error in density-sensitive calculations
improves the result substantially.
Rigorously, a precise calculation of
$\Delta E\d$ by Eq.~\ref{eq:FD} requires highly accurate densities from, e.g., correlated wavefunction calculations. 
This can only be done for smaller systems.  Moreover, the well-known difficulties of performing KS inversions
within atom-centered bases limit the precision of such calculations\cite{ZMP94,WY03}, 
which must be accurate enough to measure
the difference in density-driven error between the SC-DFT 
density and the more accurate one.\cite{NSSB20}

By definition, the theory of DC-DFT tells us that, for the vast majority of modern KS-DFT calculations,
the SC density is a very good approximation to the exact density.  
It even provides an
extremely meaningful metric of the quality of a density, 
namely how much of the energy error is due
to the density error.\cite{SSB18} 
Given the formal and practical advantages of SC densities,
we should only consider using a different density when the
density-driven error is significant.

Finally, we mention that there is extensive 
theory behind Eq.~\ref{eq:FD}.  
Ref.~\cite{VSKS19}
shows the functional analysis  that can be applied, and derives 79 more
equations from Eq.~\ref{eq:FD}.
An almost identical analysis applied
to geometries in computational chemistry (ranging from force fields
to ab initio quantum chemical methods),  produces a cornucopia of results.\cite{VB20}
The performance of such methods for molecular geometries
typically differs drastically from their energetic performance.

\subsection{HF-DFT as a DC-DFT procedure}
\label{sec_hf}

We define HF-DFT as the practice of evaluating an XC approximation on the HF density and orbitals
in all cases.\cite{tccl}
For many standard XC approximations, molecules, and properties,
when the calculation is density sensitive, HF-DFT yields more accurate results than SC-DFT.
For insensitive calculations, this may or may not be true, but it is irrelevant, as the density-driven errors are negligible.
Thus the theory of DC-DFT provides criteria for when HF-DFT should yield a better result.
The first is that the calculation should be density sensitive.   The second is that the HF density
should be a good proxy for the exact density, and hence not be, e.g., spin-contaminated
or flawed in some other way.

Thus HF-DFT is a {\em procedure}, whereas DC-DFT 
is a theoretical framework, and the two terms should not be used
interchangeably.   
We use the acronym DC(HF)-DFT to denote the use of HF-DFT {\em only} when 
a calculation is density-sensitive and its HF wavefunction is not spin-contaminated.
Unfortunately, this distinction has not always been emphasized in the past.
In their paper, Ref.~\cite{SM21} applied the HF-DFT procedure to {\em every} calculation in the
GMTKN55 database\cite{GHBE17}, which
goes against the basic principles of DC-DFT, and
provides little or no information about the performance
of DC(HF)-DFT.

By the use of the exact $n(\textbf{r})$, $\Delta E\d$ is completely eliminated (Eq.~\ref{eq:FD}), 
but $n(\textbf{r})$ is prohibitively expensive in practice.
HF densities offer a viable alternative  for reducing large density-driven errors.  In this way, only for density-sensitive calculations,
the $\tilde E[ n] \approx \tilde E[ n^{\rm HF}]$ approximation is made. 
Using $n^{\rm HF}(\textbf{r})$ in place of SC $\tilde{n}(\textbf{r})$ to reduce density-driven errors makes sense when: 
\ben
D^{HF}=\left | \tilde E[ n^{\rm HF}] - \tilde E[ n] \right |  << \left| \Delta E\d\right |.
\label{eq:D2}
\een
This condition is sufficient,
but is not always neccessary.
When HF overlocalizes,
HF-DFT can outperform SC-DFT
for functionals that delocalize,
even if Eq.~\ref{eq:D2} is not satisfied.
From the KS inversion that gives us access to highly accurate densities and KS orbitals for small systems, it has been found that Eq.~\ref{eq:D2} holds
for typical cases when $\Delta E\d$ is large.\cite{NSSB20} 
This, in turn, justifies the choice of using $n^{\rm HF}$ in place of $n$ in those calculations. 
It was also found that, for orbital-dependent XC approximations, the replacement of the KS kinetic energy with the HF kinetic energy also introduces errors
that are far smaller than $\Delta E\d$.\cite{NSSB20}

Since measuring $\Delta E\d$ by Eq.~\ref{eq:FD} requires exact densities, the more practical {\em density sensitivity} measure has been introduced:\cite{SSB18,KSSB18,NSSB20}
\ben
\tilde S= \left | \tilde E [n^{\sss LDA}]-\tilde E [n^{\sss HF}] \right|,
\label{eq:s}
\een
where tilde indicates a given functional approximation.  
This measure requires 
two non-emprirical densities: HF densities which are typically overlocalized and LDA densities which are typically delocalized.
Neither density is too expensive in routine molecular calculations.
In this way,  $\tilde S$ serves as a practical indicator of  the density sensitivity of a given reaction and
approximate functional.  
For small molecules, $\tilde S > $ 2 kcal/mol implies
density sensitivity, that the calculation may suffer from a large $\Delta E\d$, and 
for semilocal approximations, 
HF-DFT will likely improve a functional's performance.\cite{SSB18}
Therefore, SC densities should be replaced by HF ones only in density-sensitive calculations. 
Even in those cases, one has to keep an eye on whether $n^{\rm HF}$ is severely flawed by spin-contamination, as
then the use of HF densities (at least those from the spin unrestricted formalism) is not recommended either.
There is a need to develop a clever way 
to deal with the spin contamination issue, 
but for now, 
we are converting HF density to SC density for practical reasons.
The cutoff of 2 kcal/mol is taken as typical for covalent bonds
in small molecules.
As molecules grow, the cutoff must grow.\cite{MH21}
For reactions where small energy changes matters
like non-covalent interactions,
a smaller cutoff is appropriate.

\subsection{History: HF-DFT before and after DC-DFT}

The use of the HF-DFT procedure long predates the analysis of DC-DFT~\cite{GJPF92b,S92,JS08}.
As far back as the 1970's, 
Colle and Salvetti developed approximate correlation functionals 
that were trained and applied to HF densities\cite{CS75},
and led to the LYP functional\cite{LYP88}.
Other functionals, such as those of Gordon and Kim\cite{GK72}, and Wilson and Levy\cite{WL90} were
developed in a similar fashion, and these functionals were at the time highly competitive for weak interactions\cite{GK72,W05}
Even the B88 GGA\cite{B88} was developed on HF atomic densities.
Gill {\it et al.} used HF-DFT for practical reasons to test the accuracy of GGAs and hybrids without a need to obtain SC densities and orbitals for each of the functionals.\cite{GJPF92, GJPF92b}
Another reason why HF-DFT can be considered more practical than SC-DFT is that 
the former requires no grids for the XC potential at each self-consistent field (SCF) iteration.\cite{VPB12}
In all these cases, we would now say that the authors were assuming (correctly) that the calculations were density-insensitive,
and the error was dominated by the error in the XC approximation, regardless of HF versus SC density.

However, some pioneering efforts {\em did} notice that, for some problems where SC-DFT was underperforming, HF-DFT
gave systematic improvements.
This was most often observed for barrier height calculations, as demonstrated by different authors.\cite{S93,JS08,VPB12,SSB18}
We now understand that this entire class of calculation is typically density-sensitive (see below) and DC-DFT tells us why 
HF-DFT works better here but not,  say,  for main-group reaction energies.
Some of our own earlier works on electron affinities\cite{KSB11,LFB10,LB10} are couched in terms of self-interaction error and were not
understood in terms of the general principles of DC-DFT until later.\cite{WNJK17,KSB13}
On the other hand, DC-DFT has reignited interest in functionals of the HF density.\cite{GL94}
For example, it has been shown that the large-coupling strength expansion of the Møller-Plesset
adiabatic connection is a functional of the HF density.\cite{Detal20,SGVF18,DFDG21}
This finding motivated the development  of a new class of functionals applied to HF densities, which have been found
to be highly accurate for weak interactions.\cite{DFDG21}

After DC-DFT was developed\cite{KSB13}, 
explaining the success and validity of HF-DFT results,
Sim and co-workers could then identify specific classes of calculations that
are prone to density-driven errors, where HF-DFT should (and does) give improved results over SC-DFT.  
In addition to barrier heights, these include: dissociation of stretched molecules\cite{KPSS15}, some radical reaction energies\cite{KSB14}, 
electron affinities\cite{LB10,LFB10,KSB11}, specific torsional barriers\cite{NCSB21}, 
some weak interactions\cite{KSSB18}, spin gaps of Fe(II) complexes\cite{SKSB18}, 
interaction of water clusters\cite{LDPS21},
etc. 
The work of Ref.~\cite{SM21} is perhaps the most systematic benchmark of the HF-DFT procedure, but uses nothing from DC-DFT.
In the rest of this paper, 
we shall analyze their results within DC-DFT, the modern theory behind HF-DFT.\\

\begin{figure*}[htb]
\centering
\includegraphics[width=1\columnwidth]{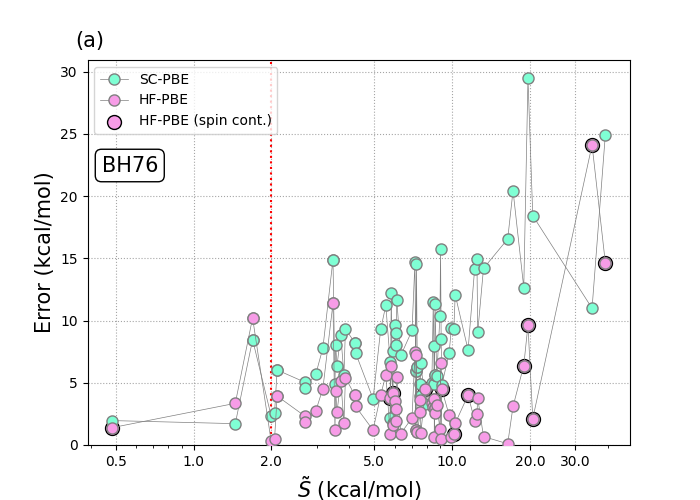}
\includegraphics[width=1\columnwidth]{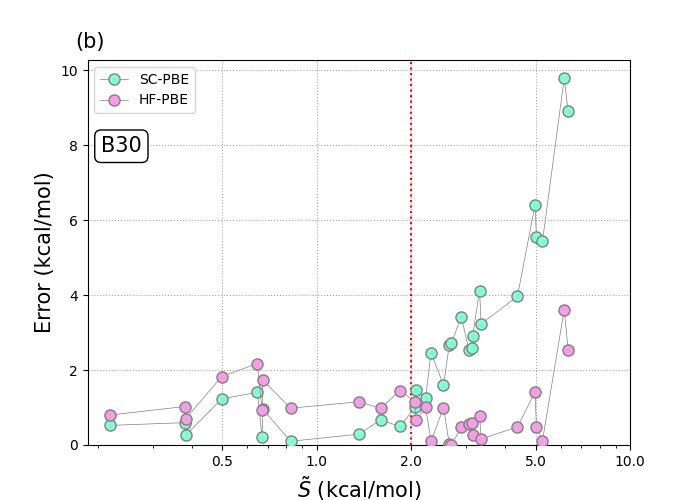}
\includegraphics[width=1\columnwidth]{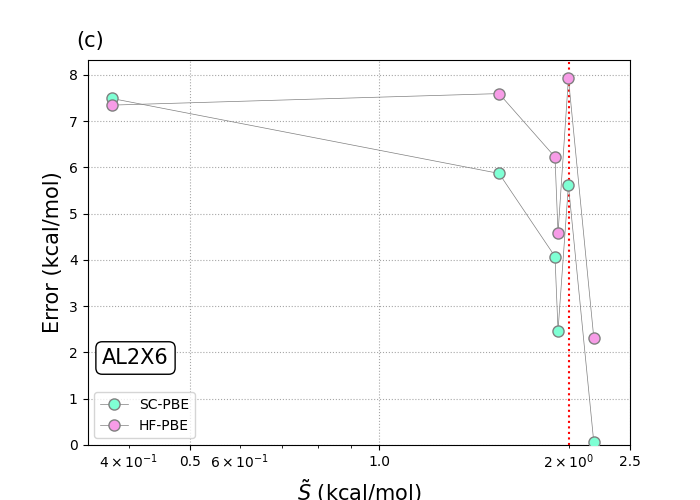}
\includegraphics[width=1\columnwidth]{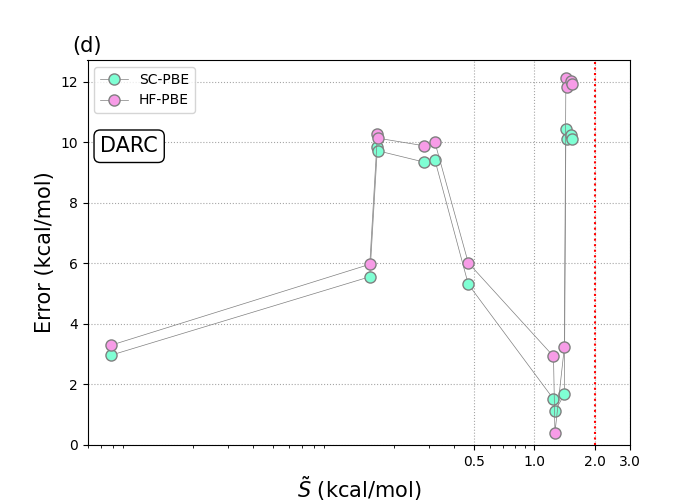}
\captionof{figure}[PBE and HF-PBE results for BH76]{
Importance of density sensitivity: Self-consistent and HF absolute errors versus density-sensitivity [Eq.~\ref{eq:s}] for PBE on the 
(a) BH76\cite{GHBE17} dataset, containing 76 hydrogen and non-hydrogen transfer
reaction barrier heights
(spin contaminated cases are marked as black marker edge color),
(b) B30\cite{BAFE13} dataset, containing 30 halogen, chalcogen and pnictogen bonding energies,
(c) AL2X6\cite{GHBE17,JMCY08} dataset, containing 6 dimerization energies
of alane derivatives,
and (d) DARC\cite{GHBE17,JMCY08} dataset, containing 14 Diels-Alder reaction energies.
}
\label{fgr:s}
\end{figure*}

\section{Specific examples}
\label{sec_ex}

\subsection{When should HF-DFT give improved results, and when should it not? 
\textit{HF-DFT works only for density sensitive cases.}}
\label{sec_sens}

The theory behind DC-DFT shows that correcting the density is relevant only when 
the self-consistent density is unusually inaccurate.   We use density sensitivity
as a proxy for this criterion and use a typical number (2 kcal/mol) as a heuristic for
small molecules with chemical bonds.   

It is not possible to see improvements by simply reporting HF-DFT numbers from large databases.
In many classes of calculations, 
the insensitive cases heavily outnumber the sensitive cases.  The gains from
using HF densities in sensitive cases can be easily hidden by the many insensitive cases,
where it often {\em worsens} energetics relative to self-consistency.

We illustrate these points with PBE calculations on
four specific databases in Fig.~\ref{fgr:s}.  The first is the BH76 database
of barrier heights.  Clearly, HF densities yield better energies when $\tilde S$ is above 2 kcal/mol,
which is the vast majority of cases in this database.  But for the few insensitive cases, we
see HF-DFT usually yields worse results.   The same pattern is repeated for B30, a database of
halogen and other bonding energies.  On the other hand, for AL2X6, most are not density-sensitive
and HF-DFT worsens the energetics.  (The one larger than 2 kcal/mol is still better self-consistently,
suggesting a higher threshold is needed here.)  
Finally, in the DARC data set, no case has
a sensitivity greater than 2 kcal/mol and self-consistent calculations are better in almost
every case.  
For this data set, HF-DFT worsens the MAE of SC-DFT by about 1 kcal/mol, 
but DC(HF)-DFT does not.

These observations illustrate the importance of measuring density errors using energies.
The errors of semilocal functionals for both Diels-Alder reactions and
barrier heights are both due to {\em delocalization error}.\cite{JMCY08,HH18}
The density errors are energetically very
relevant to barrier height energetics, but irrelevant to Diels-Alder
reactions.  DC-DFT accounts for this distinction, whereas HF-DFT does not.

\subsection{Is it easy to spot density-sensitivity?
\textit{Not always.}}
\label{sec_easy}

\begin{table*}[htb]
\centering
\resizebox{0.7\textwidth}{!}{
\begin{tabular}{llc|cc|cc|c|c}
\hline
\multicolumn{1}{l}{}   &           & $\tilde{S}_{avg}$   & SC           & HF   & SC-D4        & HF-D4        & DC(HF)                & DC(HF)-D4                                      \\ \hline
\multirow{2}{*}{BH76}       & all	& 8.0	& 8.8 	& 3.9 	& 9.3	& 4.0	& 4.4	& 4.7 \\
& w/o 12 spinc	& 6.6	& 8.5 	& 3.3 	& 8.9	& 3.6	& 3.3	& 3.5 \\
\hline
\multirow{2}{*}{RC21} &  all	& 9.2	& 5.4 	& 4.6 	& 6.9	& 4.0	& 4.3	& 4.1 \\
& w/o 9 spinc	& 11.3	& 6.8 	& 4.8 	& 8.5	& 3.8	& 4.8	& 3.8 \\
\hline
\multirow{2}{*}{RSE43} & all	& 3.7	& 3.1 	& 2.0 	& 3.0	& 2.0	& 2.0	& 1.9 \\
& w/o 8 spinc	& 2.0	& 2.9 	& 1.0 	& 2.8	& 0.9	& 1.5	& 1.5 \\
\hline
\end{tabular}
}
\caption{
PBE mean absolute errors (MAE)
on three datasets (BH76, RC21, and RSE43) 
self-consistently, with the HF density, and DC(HF).
$\tilde{S}_{avg}$ is the averaged density sensitivity for each dataset.
The deviation of HF's $\langle S^2 \rangle$ by more than 10\% from the ideal value
the criterion for spin-contamination \cite{ADLM20}.
For all 3 datasets, MAE/Root-mean-squared-displacement of absolute errors (RMSD)
values of SC-PBE becomes (5.8/4.9 kcal/mol) to (5.7/4.3) when
eliminating the spin-contaminated cases
while HF-PBE becomes (3.1/3.6) to (2.4/2.5).
The D4 parameters uses the same parameter as PBE-D4 of Ref.~\cite{CEHN19}.
See Table~\ref{tbl:spinc_all} for other RMSD information.
}
\label{tbl:spinc}
\end{table*}

In Table~\ref{tbl:spinc},  we show PBE mean absolute errors,
in its HF-,  DC(HF)-,  and SC- versions with and without the D4 dispersion enhancement\cite{CEHN19}, for
three density-sensitive data sets, including and excluding cases where   
unrestricted HF (UHF) densities are spin-contaminated.
For all three datasets,  HF-PBE improves over SC-PBE, often substantially.   On the other hand,
addition of D4 to self-consistent results often worsens them, because the D4 corrections cannot
account for poor densities.   But D4 corrections on HF densities generally do not worsen them.
Situations where Grimme's dispersion enhancement worsens self-consistent results 
due to large density-driven errors have been already described.\cite{SVSB21,MFWG21}

Next, we consider the effect of spin contamination in the UHF calculations.
For spin-contaminated cases,  DC(HF)-PBE reverts to SC-PBE, 
simply because we no longer believe
the HF density is a better approximation to the exact density.
When HF spin-contaminated
cases are excluded,  the accuracy of DC(HF)-PBE for BH76 is about the same as that of HF-PBE.  
This indicates that for BH76
even spin-contaminated HF densities
lower large PBE density-driven errors. 
But,  for RC21, 
DC(HF)-PBE improves over HF-PBE and there we can see 
the benefit of reverting DC(HF)-PBE to SC-PBE when HF is spin-contaminated.
In the case of RSE43,  the MAE of DC(HF)-PBE is about the same as that of HF-PBE. 
But when HF spin-contaminated cases are excluded,  DC(HF)-PBE worsens HF-PBE.  
Why is that? It appears that our generic cut-off of 2 kcal/mol
for density-sensitive calculations is too large for RSE43,  whose averaged absolute
energies are several times smaller than those of RC21 and BH76~\cite{GHBE17}.  
When this cut-off is lowered to 1 kcal/mol,  the MAE of DC(HF)-PBE
for RSE43 (HF spin-contaminated cases excluded) is lowered by
0.5 kcal/mol (see the right panel of Fig.~\ref{fgr:rse43}).
This demonstrates the cutoff of 2 kcal/mol
is simply a heuristic aimed at covalent bonds
in small molecules, and should be sensibly adjusted in different situations.

\begin{figure}[htb]
\centering
\includegraphics[width=0.95\columnwidth]{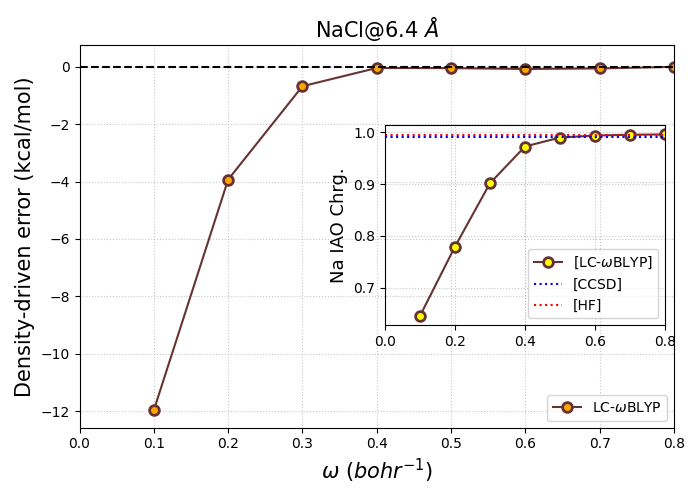}
\captionof{figure}[NaCl long-range toy model dde]{
Density-driven error ($\Delta E_D$) 
for atomization energy of NaCl at 6.4 $\AA$ bond length from LC-$\omega$BLYP 
with respect to
$\omega$ (the range-separation parameter). 
Accurate CCSD/def2-QZVPPD density and orbitals are obtained from a Kohn-Sham inversion as in Ref.~\cite{NSSB20}.
The inset shows the intrinsic atomic orbitals (IAO) population of the Na atom~\cite{K13iao}.
}
\label{fgr:na}
\end{figure}

\subsection{Do range-separated hybrids have smaller density-driven errors than their global counterparts?
{\textit Yes.}}

Ref.~\cite{SM21} found that "range-separated hybrids do not benefit much from HF-DFT
as so much HF exchange is already present at a long range." In fact, in an earlier study,
we showed
that indeed range-separated hybrids (RSHs) suffer much less from density-driven errors
than their global counterparts.\cite{SVSB21}  
For example,  semilocal DFT caclulations
of the WATER27 clusters are prone to large density-driven errors as suggested by their
large density sensitivities,  which in the case of PBE approaches 30 kcal/mol. 
Unless this density-driven error is corrected (by,  e.g.,  HF-DFT),  dispersion
corrections can even worsen the results.\cite{SVSB21,MFWG21}  In contrast to 
standard semilocal calculations of the WATER27 clusters,  a state-of-the-art
RSH developed by Mardirossian and Head-Gordon\cite{MH16}, 
$\omega$B97M-V,
suffers far less from density-driven errors,  
and its density-sensitivities for  WATER27 cluster are about 
10 times smaller than those
of PBE and 
3 times smaller those
of B3LYP\cite{SVSB21}. 
Thus DC-DFT explains why such functionals have small density-driven errors.

\begin{figure*}[htb]
\centering
\includegraphics[width=2\columnwidth]{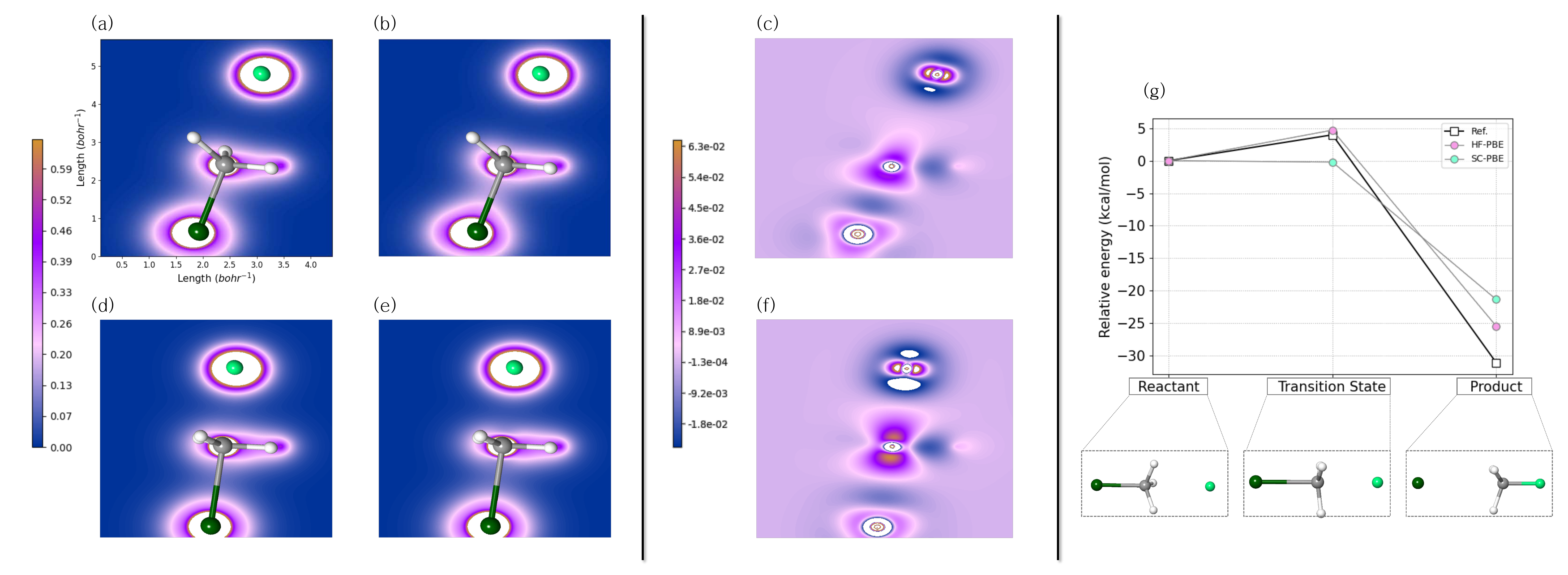}
\captionof{figure}[density_diff]{
Density plots for CH$_3$Cl+F$^-$ $\rightarrow$ FCH$_3$+Cl$^-$
(electrons per bohr$^3$):
(a) and (b) are the PBE and HF densities for 
CH$_3$Cl+F$^-$,
(d) and (e) are
their analogs for the transition state geometries, and (e) and (f) are density
differences. There is no way to deduce which method gives less density
error or which geometry is more density sensitve by just
looking at the real space density plot. But the reaction pathway plot in (g)
shows that the HF density does better than the SC density,
which was predictable from the large $S$
values for both forward
(4.9 kcal/mol) and backward (9.1 kcal/mol) barrier height.
}
\label{fgr:den_diff}
\end{figure*}

To further understand density-driven errors in RSHs, consider NaCl.
Standard semilocal functionals fail beyond about 5.8 \AA, due to incorrect
charge transfer, a specific type of delocalization error\cite{RPCV06}.
We perform calculations for NaCl at 6.4 \r{A} using LC-$\omega$BLYP,  which
employs $\omega$B88 exchange at short distances and HF exchange at long distances.
We calculate $\Delta E\d$ for the atomization energy of this system as a function of the
range-separation parameter,  $\omega$,  by using accurate density from CCSD
as done in Ref.~\cite{NSSB20}.  The results are shown in Fig.~\ref{fgr:na}.  
The inset shows the intrinsic atomic orbital (IAO)\cite{K13iao} population of the
Na atom as a function of $\omega$ and the dotted horizontal lines represents
accurate IAO from CCSD (calculated via inverted KS orbitals\cite{NSSB20}).  
The CCSD and HF's IAOs are virtually indistinguishable.  
If $\omega=0.2$ for example, the turnover distance to full exchange is sufficiently large
that significant density-driven errors would occur, just as in global hybrids.
But already at $\omega = 0.4$,  the default value used in LC-$\omega$BLYP~\cite{VydScu-JCP-06},  
$\Delta E\d$ and the deviation of the IAO charge from the reference are small.

\subsection{Can one see that the HF density is more accurate than the self-consistent
density by studying density plots?
\textit{Typically, it is hopeless.}}

There is a long history of studying density differences and density error plots in DFT, in both
physics and chemistry, with the aim of improving approximations.
Comparison is not only related to the difference between HF and self-consistent DFT approximations,
but also for the density of other methods such as MP$n$ and CASSCF.\cite{C01}
Very little insight can be gained for improving functionals from such plots, due to the
complexities of the relation between densities and potentials, as well as that between
potentials and energy functionals.  Because most good XC energy approximations yield good
densities in most calculations, despite their very poor approximate potentials, it is very
difficult to trace the effects of a good energy approximation in the (often) tiny differences
in densities.   Even in cases where we know a better density is yielding better energetics, it
is often not possible to say which aspects are relevant.  (See Fig.~4 of Ref.~\cite{SKSB18}.)

The clearest case is that of small anions.  For atomic anions, the true self-consistent
density with semilocal functionals is typically one with about 0.3 electrons escaping to 
infinity, and $Z+0.7$ electrons bound to the nucleus.  In that case, the error in the
density is glaringly obvious, even to the eye (see Fig.~2 of Ref.~\cite{KSB13}).  
But for a complicated
molecular reaction in which a small energy difference is being calculated (such as a barrier
height of the type we have already studied), the difference in such differences is typically
both small and subtle and tremendously difficult to interpret.

In Fig.~\ref{fgr:den_diff}, we illustrate this with two density difference plots,
being differences between HF and PBE densities.   These differences tend to be invisible
on the scale of the densities themselves (without labels, could one guess which density is
HF and which is PBE?) and the differences display many changes of sign.  
The upper panel (c) is for the reactants, the lower for the transition state geometry.
How can one tell that one case is density sensitive, and the other not?  
But the extreme right-hand panel tells us that there is a significant energetic
difference between the two different densities when calculating the reaction barrier,
and indeed the density sensitivity of the PBE barrier is higher than 2 kcal/mol
(4.9 kcal/mol and 9.1 kcal/mol for forward and backward each).
Moreover, PBE evaluated on the SC density predicts that the forward reactions has no barrier, whereas the PBE barrier calculated from the HF density is highly accurate. Again, this is something that one would not be able to deduce by comparing the (d) and (e) density plots.

Ref.~\cite{SM21} comments: "What is the effect of the HF density here really?
We attempted to create a difference density plot between HF-PBE and PBE for the water dimer,
but nothing of note is easily visible."  This is typically the case, but only emphasizes the importance of
the DC(HF)-DFT energtic criteria for the accuracy of densities.
In Ref.~\cite{SSB18},  handmade measures of errors in the density are too arbitrary to
be of universal use.  

DC-DFT builds an energetic criterion for measuring errors in approximate densities:
\ben
\Delta E^{ideal} [\tilde n] = E[\tilde n] - E[n],
\label{eq:ideal}
\een
which is always a single nonegative energy and provides an unambiguous measure
for all systems and approximate densities.  
In general,  obtaining $\Delta E^{ideal} [\tilde n] $ is expensive as computing $E[\tilde n]$ requires the inversion of a many-body problem and becomes more difficult than computing $E[n]$ itself.  $\Delta E^{ideal} [\tilde n] $ is typically very close to $\Delta E\d$\cite{VSKS19},  which although still expensive can be computed for more systems to gain insights into the density errors.  For larger systems,  where computing $\Delta E\d$ becomes intractable,  the sensitivity quantity of Eq.~\ref{eq:s} designed to signal large density-driven errors can be used instead.

\subsection{Can double hybrids benefit from HF-DFT?
\textit{Yes, if DC-DFT is applied very carefully.}
}

\begin{figure}[htb]
\centering
\includegraphics[width=0.95\columnwidth]{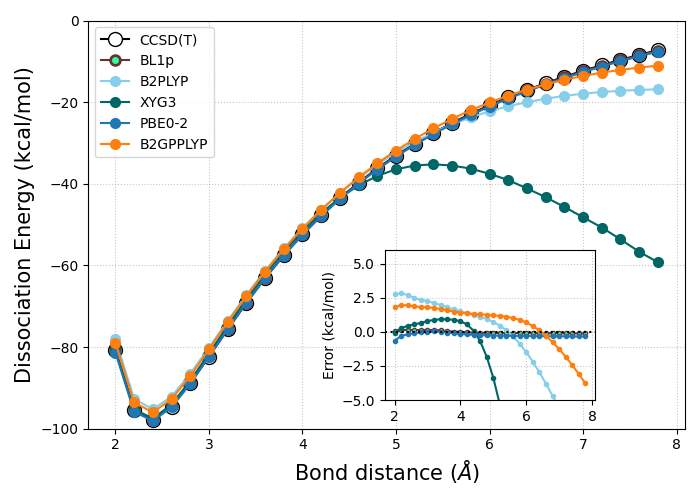}
\captionof{figure}[NaCl for DH]{
Dissociation energy curve for NaCl with various approaches: standard DHs (B2PLYP,  XYG3, PBE0-2,  B2GPPLYP),  HF-DH (BL1p) and CCSD(T) as a reference.  def2-QZVPPD basis set is used in all calculations.
Errors relative  to CCSD(T) are shown in the inset.}
\label{fgr:nacl_dh}
\end{figure}

By generalizing their findings for standard hybrids,  Ref.~\cite{SM21} concluded that HF-DFT is not beneficial for double hybrids (DHs) as these use large percentages of the HF exchange (typically $\geq 1/2$).   We agree with Ref.~\cite{SM21} that it is reasonable to expect that the higher the exact percentage in a {\em global} hybrid is,  the smaller DC-DFT correction is made by the use of HF densities.  However,  one has to be careful with the generalization of this statement to DHs given that the MP2-like energy expression evaluated on the KS orbitals is often very different from the same energy expression evaluated on the HF orbitals.\cite{CMY11}  
For that reason, the accuracy of standard DHs deteriorates when HF orbitals are used 
and so does the accuracy of HF-DHs when applied to the DFT orbitals (this is illustrated in Fig.  S4 of Ref.~\cite{SVSB21}).   At a more formal level,  the adiabatic connection formalism that underpins the construction of the recently proposed HF-DHs\cite{Detal20,DFDG21,VFGB20} is different from the usual DFT adiabatic connection used to rationalize and build standard DHs\cite{LP75,STS11}.   

Finally, we have recently developed 'BL1p',  a single-parameter HF-DH trained and
applied to HF densities.  Using principles of DC-DFT in its design,  BL1p fixes the problems of conventional
DHs  when applied to prototypical calculations suffering from large density-driven
errors.\cite{SVSB21}  
In Fig.~\ref{fgr:nacl_dh},  we show the NaCl dissociation as
a prototypical case where density-driven errors are large,
where the standard DHs fail
at large distances, despite containing large amounts of the exact exchange
but BL1p does not. 
Despite the claims of Ref.~\cite{SM21}, DC-DFT had already provided improvements in DHs\cite{SVSB21}.

\subsection{Does HF-DFT work better than self-consistent DFT for electron affinities?
\textit{Yes, because standard calculations are unconverged, as demonstrated by their positive HOMO
energies.}
}

\begin{figure}[htb]
\centering
\includegraphics[width=0.95\columnwidth]{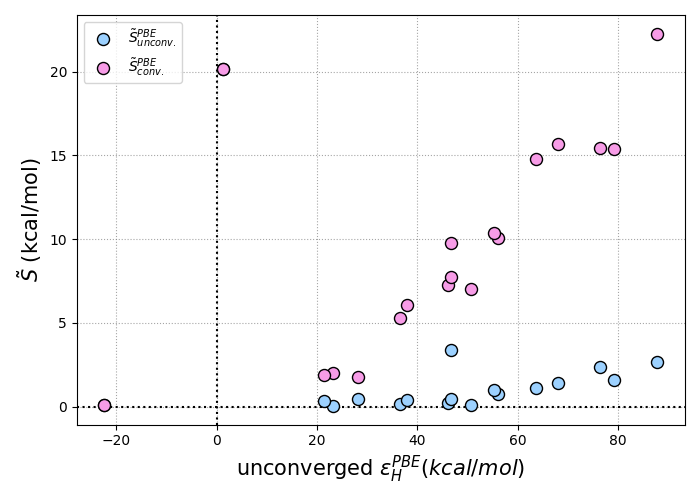}
\captionof{figure}[HOMO vs. Sensitivity]{
Relationship between unconverged HOMO energy ($\epsilon_{H}^{PBE}$) of the anion 
and density sensitivity $\tilde{S}^{PBE}$ for the electron affinities.
$\epsilon_{H}^{PBE}$ energies are mostly positive as calculated 
from a finite basis set (if a loss of a fraction of an electron is enabled in the calculations $\epsilon_{H}^{PBE}$  would hit $0$).
The results are taken from Table~\ref{tbl:g21ea},  and include both
converged (HOMO $\approx$ 0.000 a.u.)
and unconverged $\tilde S$ values.
}
\label{fgr:evsS}
\end{figure}

\begin{table*}[htb]
\centering
\resizebox{0.65\textwidth}{!}{
\begin{tabular}{l|cc|cc|ccccc}
& \multicolumn{4}{c}{Electron Affinity (kcal/mol)}                                                         & \multicolumn{3}{c}{$\epsilon_H$ (kcal/mol)} \\ \hline
\multicolumn{1}{l|}{}                           & \multicolumn{2}{c|}{PBE}                         & \multicolumn{2}{c|}{B3LYP}   & \multicolumn{4}{l}{}   \\
\multicolumn{1}{l|}{name}                           & \multicolumn{1}{c}{SC} & \multicolumn{1}{c|}{DC} & SC & \multicolumn{1}{c|}{DC} & PBE   & B3LYP   & HF & exact  \\ \hline
CH	& 3.9	& 3.1	& -1.2	& -2.3	& 56.1	& 37.5	& -47.4	& -27.9 \\
CH$_2$	& 0.5	& -0.5	& -0.8	& -1.3	& 55.2	& 37.3	& -31.7	& -13.4 \\
CH$_3$	& -2.4	& -3.4	& -5.1	& -5.7	& 63.6	& 45.2	& -18.8	& -1.2 \\
NH	& -0.3	& -1.8	& -3.0	& -4.1	& 79.3	& 57.8	& -9.3	& -8.3 \\
NH$_2$	& -0.8	& -2.0	& -4.2	& -5.1	& 68.0	& 46.4	& -29.2	& -16.8 \\
OH	& 4.1	& 0.8	& -1.0	& -3.0	& 46.7	& 22.3	& -68.7	& -41.7 \\
SiH	& 0.5	& 0.2	& -4.0	& -4.8	& 36.6	& 25.6	& -34.0	& -29.3 \\
SiH$_2$	& 2.4	& 1.8	& -1.6	& -2.6	& 37.9	& 27.0	& -30.4	& -25.1 \\
SiH$_3$	& -0.7	& -0.6	& -2.4	& -2.1	& 23.1	& 8.3	& -43.5	& -31.4 \\
PH	& -3.1	& -3.2	& -3.9	& -3.8	& 50.6	& 35.1	& -18.5	& -23.5 \\
PH$_2$	& -3.7	& -3.3	& -5.2	& -4.8	& 46.1	& 30.8	& -24.7	& -28.8 \\
HS	& -0.5	& -0.7	& -2.5	& -2.5	& 21.3	& 3.9	& -59.7	& -54.2 \\
O$_2$	& -0.1	& -2.4	& 0.2	& -1.1	& 76.4	& 47.7	& -56.9	& -9.5 \\
NO	& 3.0	& 0.3	& 3.6	& 1.6	& 87.8	& 57.1	& -59.8	& 0.2 \\
PO	& 1.3	& 0.8	& 0.0	& -0.8	& 46.7	& 28.7	& -47.8	& -24.9 \\
S$_2$	& -2.4	& -2.0	& -2.0	& -1.6	& 28.2	& 10.4	& -52.9	& -38.0 \\
Cl$_2$	& 4.6	& 4.5	& 7.6	& 7.1	& -22.4	& -45.9	& -106.8	& -54.7 \\
\hline
MAE  & 2.0 (1.5)	& 1.9 (1.3)	& 2.8 (2.0)	& 3.2 (1.8)
                         &                                         &                                         \\
MAE$^*$  & 2.9 (2.2)	& 1.8 (1.6)	& 2.1 (2.0)	& 2.0 (1.7)
                         &                                         &                                         \\
\hline \hline
\end{tabular}
}
\caption{ 
Electron affinity errors and anion HOMO energies for the G21EA dataset, omitting
CN due to large spin-contamination in HF. 
Errors relative to GMTKN95 reference, and exact HOMO is negative of electron affinity.
MAE is the mean absolute error with
RMSD values in paraenthesis.
All results 
use def2-QZVPPD basis set, except the last line, which are summary statistics using
aug-cc-pVQZ.}

\label{tbl:g21ea}
\end{table*}

It has long been known\cite{GK72} that semilocal functionals fail to bind an additional
electron to neutral atoms when a fully converged self-consistent solution is found.   This is because
the approximate KS potential (evaluated on the exact density)
is far too shallow and becomes positive not far from the nucleus and stays that way until extremely
far away (about 15 \AA).   This feature that makes it possible to extract accurate anion energies
with reasonable standard atom-centered basis, because such basis cannot represent the regions distant
from the nucleus.  Thus such calculations yield a useful representation of the density, but at the
price of having a positive HOMO, a definitive sign of an unconverged result.
Such calculations are formally unjustified, as pointed out by
R{\"o}sch and Trickey.\cite{RT97}   
In a fully numerical calculation, one finds that the minimizing
density loses about 0.3 electron to the continuum, and only binds about 1.7 electrons (for H$^-$).
Thus the truly converged self-consistent solution\cite{JD99} is extremely density sensitive
(perhaps the most visible example we know of) and properly converged calculations have large
density-driven errors.

Nonetheless, the trick of using a reasonable atomic basis does yield a reasonable density, despite
formally being a meta-stable state, and thus accurate energetics.   
In fact, many electron affinities (EA)
calculated in this way are more accurate than the corresponding ionization potentials.  
Ref.~\cite{SM21} finds that HF-DFT "appears to do more harm than good" when
electron-affinities are calculated, and indeed the numbers are comparable.
But DC-DFT provides formally more solid ground for calculating electron affinities than
the standard DFT approximations with limited basis sets.\cite{KSB11,LFB10}

Consider what happens when we try to add an extra electron to the hydrogen atom within the PBE functional.
First, we do the calculation in a large atomic basis.
For the electron affinity of the hydrogen atom,  PBE/aug-cc-pV6Z yields an error
that is less than 1 kcal/mol (see Table~\ref{tbl:Hea} for numerical details).  
Moreover, the density sensitivity of this calculation is 
about 1 kcal/mol,  which  is less than our sensitivity threshold. 
This consideration would naively indicate that there is no 
need to invoke HF-DFT.  But the positive HOMO of PBE reminds
us that H$^-$ is artificially bound by a finite 
atom-centered basis set despite its large size (aug-cc-pV6Z).  
The results for both electron affinity and $\tilde S$ are unconverged. 
If, instead, we converge the PBE calculation of H$^-$  by enabling fractional occupations
until the HOMO energy hits zero (as done in Ref.~\cite{SSB18}), things change dramatically.  
About 0.37 of an electron is lost confirming that
the converged PBE calculation does not even bind H$^-$.  
Moreover, the converged $\tilde S$ becomes 8 kcal/mol, showing that the calculation is 
in fact extremely density-sensitive. 

We now repeat this test for the G21EA dataset and compare
converged (a fraction of charge enabled until the HOMO hits zero)
vs.  unconverged (anions bound only by a finite basis) $\tilde S$ of PBE.
These are shown in Figure~\ref{fgr:evsS}
as a function of the HOMO energies of the anions
(these are mostly positive as calculated from a finite basis set).   
First we note that the two sensitivities are identical in only one case (Cl$_2$),
where self-consistent PBE gives a negative HOMO energy and therefore a bound anion.  In all other cases,
the unconverged calculations severely underestimate $\tilde S$.
Furthermore,   Fig.~\ref{fgr:evsS} also shows a clear correlation 
between the HOMO energy of the anion and $\tilde S$ of EA.  
In nearly all cases, the converged $\tilde S$ is above our 2 kcal/mol threshold.
Thus DC-DFT invokes HF-DFT for calculations of the affinities given that
HF HOMOs of the anions are bound and properly converged PBE calculations
in nearly all cases give unbound anions. 

What about the numbers? Basis-set bound DFT calculations 
give reasonable accuracies for the electron affinities.  
Table~\ref{tbl:g21ea} shows
electron affinity errors of PBE and B3LYP and their HF-DFT counterparts, 
and the corresponding anion HOMO energies.
The MAEs of DC-DFT and basis-set bound (unconverged) DFT are comparable.
But note that the RMSD values are noticeably smaller for DC-DFT, showing a smaller
variance and less outliers.  
Moreover, if the smaller aug-cc-pVQZ basis set is used instead, as it often is in practice,
many results shift by amounts comparable to these errors (see Table S3).
Finally, we note that these molecules form the G21EA data set, with CN excluded as its HF wavefunction
is heavily spin contaminated, so DC-DFT simply reverts to SC-DFT for that case.
If, however, restricted open-shell HF (ROHF) is used instead of UHF for the CN molecule,
the error of HF-PBE is greatly reduced.  
We currently use UHF as a default in our HF-DFT calculations of
open-shell systems and we will further investigate the usability of ROHF-DFT\cite{PBMM17} in the future.

\subsection{Is slow grid convergence as much a problem for HF-SCAN as it is for SCAN?
\textit{Yes, it can be.}
}

Ref.~\cite{SM21} have shown the  HF-SCAN displays an excellent performance for the GMTKN55 database and gets beaten only by few modern and highly parameterized functionals.  The painfully slow convergence of 
SC-SCAN calculations is well known,  
and  even for simple atomization energies  
SCAN is difficult to converge within the available grid sizes in many standard quantum-chemical codes.\cite{FKNP20}  In the case of HF-SCAN,  the orbitals are taken from a HF calculation so grids are only needed for the final energy evaluation and not for the whole SCF procedure.  For that reason,  one would assume that  the grid convergence is less of a problem for HF-SCAN.  Nevertheless,  as shown in Fig.~\ref{fgr:g21ip_grid_2} for the  G21IP dataset,  HF-SCAN is also not converging within the grid sizes available in the quantum-chemical package ORCA.  The r$^2$SCAN functional is designed to fix the grid convergence issues of SCAN,  and as shown in Fig. ~\ref{fgr:g21ip_grid_2},  the HF-r$^2$SCAN results converge even within smaller ORCA grids.  At the same time,  the accuracy of 
HF-SCAN is comparable to HF-r$^2$SCAN (see, e.g., Fig.~\ref{fgr:g21ip_grid_2} illustrating this for the G21IP dataset). 
These difficulties are overcome by using
extremely large grids in Qchem,
showing that HF-SCAN can yield
highly accurate MB-pol potentials for water simulations.\cite{DLPP21}

\subsection{What are the current limitations of DC-DFT and where else can it be applied?}

\textit{Homonuclear dimers:} As a form of DC-DFT,  HF-DFT is meant to reduce only density-driven errors and not functional errors.  In cases where DFT errors are large even when an approximate functional is evaluated on the exact density,  HF-DFT cannot help.  While HF-DFT substantially reduces the errors of DFT in stretched heterodimers,  this is not the case for homodimers.  For example,  upon stretching H$_2^+$ or  He$_2^+$,  the standard DFT methods have large self-interaction errors,  only a tiny fraction of which is density-driven.\cite{SVSB21}  
HF-DFT also cannot help stretching H$_2$\cite{KSB13},  
the case where standard DFT methods have large static correlation error\cite{CMY11}. 

\textit{Transition metals:} DC(HF)-DFT has been mostly applied to main-group chemistry,  but it also can be greatly useful in the transition-metal chemistry (see,  e.g.,  Ref.~\cite{SKSB18} for the calculations of spin gaps of Fe(II) complexes).  
Further study on the usefulness of HF-DFT 
for the transition-metal chemsitry is warranted.\cite{MH21}

\textit{Materials and surface science calculations:} Different forms of DC-DFT have been also used to analyse the errors of DFT for surface science applications.  The cases where more accurate densities improves the results of semilocal functionals drastically improves the adsorption description of CO on metallic surfaces have been identified.\cite{PPSP19}  In other cases,  such as the challenging barrier height for attaching O$_2$ to the Al(111) surface,   GGAs benefit little from more accurate densities.\cite{GSVPDK20}  This implies that the accuracy of their density is satisfactory at the GGA level and evaluating a hybrid functional on those GGA densities gives results  similar to fully SC-hybrid calculations,  but at a fraction of the cost.\cite{GSVPDK20}  Similar cost reduction strategies have been applied to semiconductor calculations.\cite{SGMP20}  In future years,  where and how DC-DFT 
can be useful
in surface and material science needs to be further explored.

\textit{Forces:} Despite being a non-SCF scheme,  the calculation of the analytical gradients for HF-DFT is just slightly more involved than those of SC-DFT.  This has been demonstrated by Bartlett and co-workers\cite{VPB12,SOB94}.  The more recent Pyxdh code\cite{pyxdh}  is written based on the PYSCF package\cite{PYSCF} and with little  modifications can be used for HF-DFT geometry optmizations.  Besides earlier calculations of barrier heights by using HF-DFT optmized transition states\cite{VPB12} and exploration of radical potential energy surfaces\cite{KSB14},  the performance of HF-DFT for molecular structures relative to their SC-DFT counterparts is yet to be systematically explored.  This is another topic that we will cover in the future,  and recently developed unambigious measures for assessing the quality of approximate molecular structures should prove useful.\cite{VB20}

\section{Conclusions}
\label{concs}

We hope this short work explains and clarifies many of the basic issues
behind density-corrected DFT.  Most important among these are (a) that
there exist many situations where the energy error in a DFT calculation
has a significant density-driven contribution, (b) that blind application
of HF-DFT (using DFT approximations on HF densities) is not DC-DFT and
most (if not all) of the benefits of DC-DFT are lost in the details of
large databases, but (c) careful application of the principles of DC-DFT
shows that DC-(HF)-DFT, as a practical scheme, almost always works for
density-sensitive cases.

SS and ES are grateful for support from the National Research Foundation of Korea (NRF-2020R1A2C2007468 and NRF-2020R1A4A1017737). 
KB acknowledges funding from NSF (CHEM 1856165).
SV acknowledges funding from the Marie Sk\l{}odowska-Curie grant 101033630 (EU’s Horizon 2020 programme).
We thanks G.Santra and J.M.L. Martin for helpful discussions and providing the data on which Ref.~\cite{SM21} was based.

\bibliographystyle{unsrt}

\begin{thebibliography}{10}

\bibitem{MH17}
Narbe Mardirossian and Martin Head-Gordon.
\newblock Thirty years of density functional theory in computational chemistry:
  an overview and extensive assessment of 200 density functionals.
\newblock {\em Molecular Physics}, 115(19):2315--2372, 2017.

\bibitem{WNJK17}
Adam Wasserman, Jonathan Nafziger, Kaili Jiang, Min-Cheol Kim, Eunji Sim, and
  Kieron Burke.
\newblock The importance of being self-consistent.
\newblock {\em Annual Review of Physical Chemistry}, 68(1):555--581, 2017.

\bibitem{KSB13}
Min-Cheol Kim, Eunji Sim, and Kieron Burke.
\newblock Understanding and reducing errors in density functional calculations.
\newblock {\em Physical Review Letters}, 111(7):073003, 2013.

\bibitem{KSB14}
Min-Cheol Kim, Eunji Sim, and Kieron Burke.
\newblock Ions in solution: Density corrected density functional theory
  (dc-dft).
\newblock {\em The Journal of Chemical Physics}, 140(18):18A528, 2014.

\bibitem{KPSS15}
Min-Cheol Kim, Hansol Park, Suyeon Son, Eunji Sim, and Kieron Burke.
\newblock Improved dft potential energy surfaces via improved densities.
\newblock {\em The Journal of Physical Chemistry Letters}, 6(19):3802--3807,
  2015.

\bibitem{KSSB18}
Yeil Kim, Suhwan Song, Eunji Sim, and Kieron Burke.
\newblock Halogen and chalcogen binding dominated by density-driven errors.
\newblock {\em The Journal of Physical Chemistry Letters}, 10(2):295--301,
  2018.

\bibitem{KSB11}
Min-Cheol Kim, Eunji Sim, and Kieron Burke.
\newblock Communication: Avoiding unbound anions in density functional
  calculations.
\newblock {\em The Journal of Chemical Physics}, 134(17):171103, 2011.

\bibitem{NSSB20}
Seungsoo Nam, Suhwan Song, Eunji Sim, and Kieron Burke.
\newblock Measuring density-driven errors using kohn{\textendash}sham
  inversion.
\newblock {\em Journal of Chemical Theory and Computation}, 16(8):5014--5023,
  July 2020.

\bibitem{SKSB18}
Suhwan Song, Min-Cheol Kim, Eunji Sim, Anouar Benali, Olle Heinonen, and Kieron
  Burke.
\newblock Benchmarks and reliable dft results for spin gaps of small ligand fe
  (ii) complexes.
\newblock {\em Journal of Chemical Theory and Computation}, 14(5):2304--2311,
  2018.

\bibitem{CC90}
Enrico Clementi and Subhas~J Chakravorty.
\newblock A comparative study of density functional models to estimate
  molecular atomization energies.
\newblock {\em The Journal of Chemical Physics}, 93(4):2591--2602, 1990.

\bibitem{GJPF92}
Peter~MW Gill, Benny~G Johnson, John~A Pople, and Michael~J Frisch.
\newblock An investigation of the performance of a hybrid of hartree-fock and
  density functional theory.
\newblock {\em International Journal of Quantum Chemistry}, 44(S26):319--331,
  1992.

\bibitem{GJPF92b}
Peter~MW Gill, Benny~G Johnson, John~A Pople, and Michael~J Frisch.
\newblock The performance of the Becke—Lee—Yang—Parr (B—LYP) density functional theory with various basis sets.
\newblock {\em Chemical Physics Letters}, 197(4-5):499--505, 1992.

\bibitem{HMAA92}
Nicholas~C Handy, Paul~E Maslen, Roger~D Amos, Jamie~S Andrews, Christopher~W
  Murray, and Gregory~J Laming.
\newblock The harmonic frequencies of benzene.
\newblock {\em Chemical physics letters}, 197(4-5):506--515, 1992.

\bibitem{S92}
Gustavo~E Scuseria.
\newblock Comparison of coupled-cluster results with a hybrid of hartree--fock
  and density functional theory.
\newblock {\em The Journal of chemical physics}, 97(10):7528--7530, 1992.

\bibitem{JGP92}
Benny~G Johnson, Peter~MW Gill, and John~A Pople.
\newblock Preliminary results on the performance of a family of density
  functional methods.
\newblock {\em The Journal of chemical physics}, 97(10):7846--7848, 1992.

\bibitem{OB94}
Nevin Oliphant and Rodney~J Bartlett.
\newblock A systematic comparison of molecular properties obtained using
  hartree--fock, a hybrid hartree--fock density-functional-theory, and
  coupled-cluster methods.
\newblock {\em The Journal of Chemical Physics}, 100(9):6550--6561, 1994.

\bibitem{J15}
R.~O. Jones.
\newblock Density functional theory: Its origins, rise to prominence, and
  future.
\newblock {\em Rev. Mod. Phys.}, 87:897--923, Aug 2015.

\bibitem{RH16}
Jan Rezac and Pavel Hobza.
\newblock Benchmark calculations of interaction energies in noncovalent
  complexes and their applications.
\newblock {\em Chemical reviews}, 116(9):5038--5071, 2016.

\bibitem{GHBE17}
Lars Goerigk, Andreas Hansen, Christoph Bauer, Stephan Ehrlich, Asim Najibi,
  and Stefan Grimme.
\newblock A look at the density functional theory zoo with the advanced gmtkn55
  database for general main group thermochemistry, kinetics and noncovalent
  interactions.
\newblock {\em Physical Chemistry Chemical Physics}, 19(48):32184--32215, 2017.

\bibitem{MH16}
Narbe Mardirossian and Martin Head-Gordon.
\newblock $\omega$ b97m-v: A combinatorially optimized, range-separated hybrid,
  meta-gga density functional with vv10 nonlocal correlation.
\newblock {\em The Journal of chemical physics}, 144(21):214110, 2016.

\bibitem{SSB18}
Eunji Sim, Suhwan Song, and Kieron Burke.
\newblock Quantifying density errors in dft.
\newblock {\em The Journal of Physical Chemistry Letters}, 9(22):6385--6392,
  2018.

\bibitem{PRCS09}
John~P. Perdew, Adrienn Ruzsinszky, Lucian~A. Constantin, Jianwei Sun, and
  Gábor~I. Csonka.
\newblock Some fundamental issues in ground-state density functional theory: A
  guide for the perplexed.
\newblock {\em Journal of Chemical Theory and Computation}, 5(4):902--908,
  2009.
\newblock PMID: 26609599.

\bibitem{SM21}
Golokesh Santra and Jan~ML Martin.
\newblock What types of chemical problems benefit from density-corrected dft? a
  probe using an extensive and chemically diverse test suite.
\newblock {\em Journal of chemical theory and computation}, 17(3):1368--1379,
  2021.

\bibitem{KS65}
Walter Kohn and Lu~Jeu Sham.
\newblock Self-consistent equations including exchange and correlation effects.
\newblock {\em Physical Review}, 140(4A):A1133, 1965.

\bibitem{ZMP94}
Qingsheng Zhao, Robert~C Morrison, and Robert~G Parr.
\newblock From electron densities to kohn-sham kinetic energies, orbital
  energies, exchange-correlation potentials, and exchange-correlation energies.
\newblock {\em Physical Review A}, 50(3):2138, 1994.

\bibitem{WY03}
Qin Wu and Weitao Yang.
\newblock A direct optimization method for calculating density functionals and
  exchange--correlation potentials from electron densities.
\newblock {\em The Journal of Chemical Physics}, 118(6):2498--2509, 2003.

\bibitem{VSKS19}
Stefan Vuckovic, Suhwan Song, John Kozlowski, Eunji Sim, and Kieron Burke.
\newblock Density functional analysis: The theory of density-corrected dft.
\newblock {\em Journal of Chemical Theory and Computation}, 15(12):6636--6646,
  2019.

\bibitem{VB20}
Stefan Vuckovic and Kieron Burke.
\newblock Quantifying and understanding errors in molecular geometries.
\newblock {\em The Journal of Physical Chemistry Letters}, 11(22):9957--9964,
  November 2020.
  
\bibitem{tccl}
Yonsei~University Theoretical and Computational~Chemistry Laboratory.
\newblock Density corrected-density functional theory.
\newblock \url{http://tccl.yonsei.ac.kr/mediawiki/index.php/DC-DFT}.


\bibitem{MH21}
Carlos Mart{\'\i}n-Fern{\'a}ndez and Jeremy~N Harvey.
\newblock On the use of normalized metrics for density sensitivity analysis in
  dft.
\newblock {\em The Journal of Physical Chemistry A}, 2021.

\bibitem{JS08}
Benjamin~G Janesko and Gustavo~E Scuseria.
\newblock Hartree--fock orbitals significantly improve the reaction barrier
  heights predicted by semilocal density functionals.
\newblock {\em The Journal of Chemical Physics}, 128(24):244112, 2008.

\bibitem{CS75}
Renato Colle and Oriano Salvetti.
\newblock Approximate calculation of the correlation energy for the closed
  shells.
\newblock {\em Theoretica Chimica Acta}, 37(4):329--334, 1975.

\bibitem{LYP88}
Chengteh Lee, Weitao Yang, and Robert~G Parr.
\newblock Development of the colle-salvetti correlation-energy formula into a
  functional of the electron density.
\newblock {\em Physical Review B}, 37(2):785--789, 1988.

\bibitem{GK72}
Roy~G Gordon and Yung~Sik Kim.
\newblock Theory for the forces between closed-shell atoms and molecules.
\newblock {\em The Journal of Chemical Physics}, 56(6):3122--3133, 1972.

\bibitem{WL90}
Leslie~C Wilson and Mel Levy.
\newblock Nonlocal wigner-like correlation-energy density functional through
  coordinate scaling.
\newblock {\em Physical Review B}, 41(18):12930, 1990.

\bibitem{W05}
Tiffany~R Walsh.
\newblock Exact exchange and wilson--levy correlation: a pragmatic device for
  studying complex weakly-bonded systems.
\newblock {\em Physical Chemistry Chemical Physics}, 7(3):443--451, 2005.

\bibitem{B88}
Axel~D Becke.
\newblock Density-functional exchange-energy approximation with correct
  asymptotic behavior.
\newblock {\em Physical Review A}, 38(6):3098--3100, 1988.

\bibitem{VPB12}
Prakash Verma, Ajith Perera, and Rodney~J. Bartlett.
\newblock Increasing the applicability of dft i: Non-variational correlation
  corrections from hartree-fock dft for predicting transition states.
\newblock {\em Chemical Physics Letters}, 524:10 -- 15, 2012.

\bibitem{S93}
Jorge~M Seminario.
\newblock Energetics using dft: comparions to precise ab initio and experiment.
\newblock {\em Chemical physics letters}, 206(5-6):547--554, 1993.

\bibitem{LFB10}
Donghyung Lee, Filipp Furche, and Kieron Burke.
\newblock Accuracy of electron affinities of atoms in approximate density
  functional theory.
\newblock {\em The Journal of Physical Chemistry Letters}, 1(14):2124--2129,
  2010.

\bibitem{LB10}
Donghyung Lee and Kieron Burke.
\newblock Finding electron affinities with approximate density functionals.
\newblock {\em Molecular Physics}, 108(19-20):2687--2701, 2010.

\bibitem{GL94}
Andreas G{\"o}rling and Mel Levy.
\newblock Exact kohn-sham scheme based on perturbation theory.
\newblock {\em Physical Review A}, 50(1):196, 1994.

\bibitem{Detal20}
Timothy~J Daas, Juri Grossi, Stefan Vuckovic, Ziad~H Musslimani, Derk~P Kooi,
  Michael Seidl, Klaas~JH Giesbertz, and Paola Gori-Giorgi.
\newblock Large coupling-strength expansion of the m{\o}ller--plesset adiabatic
  connection: From paradigmatic cases to variational expressions for the
  leading terms.
\newblock {\em The Journal of Chemical Physics}, 153(21):214112, 2020.

\bibitem{SGVF18}
Michael Seidl, Sara Giarrusso, Stefan Vuckovic, Eduardo Fabiano, and Paola
  Gori-Giorgi.
\newblock Communication: Strong-interaction limit of an adiabatic connection in
  hartree-fock theory.
\newblock {\em The Journal of chemical physics}, 149(24):241101, 2018.

\bibitem{DFDG21}
Timothy~J Daas, Eduardo Fabiano, Fabio Della~Sala, Paola Gori-Giorgi, and
  Stefan Vuckovic.
\newblock Noncovalent interactions from models for the m{\o}ller--plesset
  adiabatic connection.
\newblock {\em The journal of physical chemistry letters}, 12:4867--4875, 2021.

\bibitem{NCSB21}
Seungsoo Nam, Eunbyol Cho, Eunji Sim, and Kieron Burke.
\newblock Explaining and fixing dft failures for torsional barriers.
\newblock {\em The journal of physical chemistry letters}, 12(11):2796--2804,
  2021.

\bibitem{LDPS21}
Eleftherios Lambros, Saswata Dasgupta, Etienne Palos, Steven Swee, Jie Hu, and
  Francesco Paesani.
\newblock General many-body framework for data-driven potentials with arbitrary
  quantum mechanical accuracy: Water as a case study.
\newblock {\em Journal of Chemical Theory and Computation}, 17(9):5635--5650,
  2021.

\bibitem{BAFE13}
Antonio Bauza, Ibon Alkorta, Antonio Frontera, and Jose Elguero.
\newblock On the reliability of pure and hybrid dft methods for the evaluation
  of halogen, chalcogen, and pnicogen bonds involving anionic and neutral
  electron donors.
\newblock {\em Journal of Chemical Theory and Computation}, 9(11):5201--5210,
  2013.

\bibitem{JMCY08}
Erin~R Johnson, Paula Mori-S{\'a}nchez, Aron~J Cohen, and Weitao Yang.
\newblock Delocalization errors in density functionals and implications for
  main-group thermochemistry.
\newblock {\em The Journal of Chemical Physics}, 129(20):204112, 2008.

\bibitem{HH18}
Diptarka Hait and Martin Head-Gordon.
\newblock Delocalization errors in density functional theory are essentially
  quadratic in fractional occupation number.
\newblock {\em The journal of physical chemistry letters}, 9(21):6280--6288,
  2018.

\bibitem{ADLM20}
Adam Rettig, Diptarka Hait, Luke~W Bertels, and Martin Head-Gordon.
\newblock Third-order m{\o}ller--plesset theory made more useful? the role of
  density functional theory orbitals.
\newblock {\em Journal of Chemical Theory and Computation}, 16(12):7473--7489,
  2020.

\bibitem{CEHN19}
Eike Caldeweyher, Sebastian Ehlert, Andreas Hansen, Hagen Neugebauer, Sebastian
  Spicher, Christoph Bannwarth, and Stefan Grimme.
\newblock A generally applicable atomic-charge dependent london dispersion
  correction.
\newblock {\em The Journal of Chemical Physics}, 150(15):154122, 2019.

\bibitem{SVSB21}
Suhwan Song, Stefan Vuckovic, Eunji Sim, and Kieron Burke.
\newblock Density sensitivity of empirical functionals.
\newblock {\em The journal of physical chemistry letters}, 12(2):800--807,
  2021.

\bibitem{MFWG21}
Nisha Mehta, Thomas Fellowes, Jonathan~M White, and Lars Goerigk.
\newblock Chal336 benchmark set: How well do quantum-chemical methods describe
  chalcogen-bonding interactions?
\newblock {\em Journal of Chemical Theory and Computation}, 17(5):2783--2806,
  2021.

\bibitem{K13iao}
Gerald Knizia.
\newblock Intrinsic atomic orbitals: An unbiased bridge between quantum theory
  and chemical concepts.
\newblock {\em Journal of chemical theory and computation}, 9(11):4834--4843,
  2013.

\bibitem{RPCV06}
Adrienn Ruzsinszky, John~P Perdew, G{\'a}bor~I Csonka, Oleg~A Vydrov, and
  Gustavo~E Scuseria.
\newblock Spurious fractional charge on dissociated atoms: Pervasive and
  resilient self-interaction error of common density functionals.
\newblock {\em The Journal of Chemical Physics}, 125(19):194112, 2006.

\bibitem{VydScu-JCP-06}
O.~A. Vydrov and G.~E. Scuseria.
\newblock Assessment of a long-range corrected hybrid functional.
\newblock {\em J. Chem. Phys.}, {125}:234109, 2006.

\bibitem{C01}
Dieter Cremer.
\newblock Density functional theory: coverage of dynamic and non-dynamic
  electron correlation effects.
\newblock {\em Molecular Physics}, 99(23):1899--1940, 2001.

\bibitem{CMY11}
Aron~J Cohen, Paula Mori-S{\'a}nchez, and Weitao Yang.
\newblock Challenges for density functional theory.
\newblock {\em Chemical Reviews}, 112(1):289--320, 2011.

\bibitem{VFGB20}
Stefan Vuckovic, Eduardo Fabiano, Paola Gori-Giorgi, and Kieron Burke.
\newblock Map: An mp2 accuracy predictor for weak interactions from adiabatic
  connection theory.
\newblock {\em Journal of Chemical Theory and Computation}, 16(7):4141--4149,
  May 2020.
\newblock PMID: 32379454.

\bibitem{LP75}
David~C Langreth and John~P Perdew.
\newblock The exchange-correlation energy of a metallic surface.
\newblock {\em Solid State Communications}, 17(11):1425--1429, 1975.

\bibitem{STS11}
Kamal Sharkas, Julien Toulouse, and Andreas Savin.
\newblock Double-hybrid density-functional theory made rigorous.
\newblock {\em The Journal of Chemical Physics}, 134(6):064113, 2011.

\bibitem{RT97}
Notker R{\"o}sch and SB~Trickey.
\newblock Comment on “Concerning the applicability of density functional methods to atomic and molecular negative ions”[J. Chem. Phys. 105, 862 (1996)].
\newblock {\em The Journal of Chemical Physics}, 106(21):8940--8941, 1997.

\bibitem{JD99}
Andrzej~A Jar{\k{e}}cki and Ernest~R Davidson.
\newblock Density functional theory calculations for f-.
\newblock {\em Chemical Physics Letters}, 300(1-2):44--52, 1999.

\bibitem{PBMM17}
Robert~CR Pennifold, Simon~J Bennie, Thomas~F Miller~III, and Frederick~R
  Manby.
\newblock Correcting density-driven errors in projection-based embedding.
\newblock {\em The Journal of chemical physics}, 146(8):084113, 2017.

\bibitem{FKNP20}
James~W Furness, Aaron~D Kaplan, Jinliang Ning, John~P Perdew, and Jianwei Sun.
\newblock Accurate and numerically efficient r2scan meta-generalized gradient
  approximation.
\newblock {\em The journal of physical chemistry letters}, 11(19):8208--8215,
  2020.

\bibitem{DLPP21}
Saswata Dasgupta, Eleftherios Lambros, John Perdew, and Francesco Paesani.
\newblock Elevating density functional theory to chemical accuracy for water
  simulations through a density-corrected many-body formalism.
\newblock {\em ChemRxiv doi:{\tt 10.33774/chemrxiv-2021-hstgf-v3}}, 2021.

\bibitem{PPSP19}
Abhirup Patra, Haowei Peng, Jianwei Sun, and John~P. Perdew.
\newblock Rethinking {CO} adsorption on transition-metal surfaces: Effect of
  density-driven self-interaction errors.
\newblock {\em Physical Review B}, 100(3), July 2019.

\bibitem{GSVPDK20}
Nick Gerrits, Egidius W.~F. Smeets, Stefan Vuckovic, Andrew~D. Powell,
  Katharina Doblhoff-Dier, and Geert-Jan Kroes.
\newblock Density functional theory for molecule{\textendash}metal surface
  reactions: When does the generalized gradient approximation get it right, and
  what to do if it does not.
\newblock {\em The Journal of Physical Chemistry Letters}, 11(24):10552--10560,
  December 2020.

\bibitem{SGMP20}
Jonathan~M. Skelton, David S.~D. Gunn, Sebastian Metz, and Stephen~C. Parker.
\newblock Accuracy of hybrid functionals with non-self-consistent
  kohn{\textendash}sham orbitals for predicting the properties of
  semiconductors.
\newblock {\em Journal of Chemical Theory and Computation}, 16(6):3543--3557,
  May 2020.

\bibitem{SOB94}
Hideo Sekino, Nevin Oliphant, and Rodney~J Bartlett.
\newblock Property evaluation using the hartree--fock-density-functional-theory
  method: An efficient formalism for first-and second-order properties.
\newblock {\em The Journal of chemical physics}, 101(9):7788--7794, 1994.

\bibitem{pyxdh}
Python xdh project, available from {\tt https://github.com/ajz34/py\_xdh}.

\bibitem{PYSCF}
Qiming Sun, Timothy~C. Berkelbach, Nick~S. Blunt, George~H. Booth, Sheng Guo,
  Zhendong Li, Junzi Liu, James~D. McClain, Elvira~R. Sayfutyarova, Sandeep
  Sharma, Sebastian Wouters, and Garnet Kin‐Lic Chan.
\newblock PySCF: the Python‐based simulations of chemistry framework, 2017.

\end{thebibliography}

\newcommand{\beginsupplement}{%
        \setcounter{table}{0}
        \renewcommand{\thetable}{S\arabic{table}}%
        \setcounter{figure}{0}
        \renewcommand{\thefigure}{S\arabic{figure}}%
     }

\beginsupplement

\clearpage

\section{Supporting Information}

\begin{table}[htb]
\centering
\begin{tabular}{lll}
\hline
CCSD              & 17.3 &  \\
HF-PBE            & 15.6 &  \\
PBE (un-conv.)    & 16.6 &  \\
PBE (conv.)       & 23.9 &  \\ \hline
$S^{PBE}$ (un-conv.) & 1.1  &  \\
$S^{PBE}$ (conv.)    & 8.4  &  \\
\end{tabular}
\caption{
Electron affinity information of hydrogen atom. CCSD/aug-cc-pV6Z is used as a reference. For standard PBE functional, 0.37 electrons are unbound and denoted as un-converged. For converged cases, where electrons are omitted to match HOMO equals 0, denoted as converged.
}
\label{tbl:Hea}
\end{table}

\begin{table*}[htb]
\centering
\resizebox{0.9\textwidth}{!}{
\begin{tabular}{llccccccc}
\hline
\multicolumn{1}{l}{}   &           & $\tilde{S}_{avg.}$   & SC           & HF           & DC(HF)        & SC-D4        & HF-D4        & DC(HF)-D4                                      \\ \hline
\multirow{2}{*}{BH76}       & 12 spinc	& 8.0	& 8.8 (5.0)	& 3.9 (3.7)	& 4.4 (5.0)	& 9.3 (5.1)	& 4.0 (3.7)	& 4.7 (5.1) \\
& w/o spinc	& 6.6	& 8.5 (3.8)	& 3.3 (2.6)	& 3.3 (2.4)	& 8.9 (3.9)	& 3.6 (2.7)	& 3.5 (2.6) \\
\multirow{2}{*}{RC21} & 9 spinc	& 9.2	& 5.4 (3.2)	& 4.6 (3.5)	& 4.3 (3.1)	& 6.9 (3.7)	& 4.0 (2.7)	& 4.1 (2.9) \\
& w/o spinc	& 11.3	& 6.8 (2.8)	& 4.8 (3.2)	& 4.8 (3.2)	& 8.5 (2.9)	& 3.8 (2.4)	& 3.8 (2.4) \\
\multirow{2}{*}{RSE43} & 8 spinc	& 3.7	& 3.1 (1.4)	& 2.0 (3.3)	& 2.0 (1.7)	& 3.0 (1.4)	& 2.0 (3.4)	& 1.9 (1.6) \\
& w/o spinc	& 2.0	& 2.9 (1.2)	& 1.0 (0.8)	& 1.5 (1.1)	& 2.8 (1.2)	& 0.9 (0.7)	& 1.5 (1.0) \\
\hline
\hline
\end{tabular}
}
\caption{
PBE mean absolute errors (MAE, kcal/mol) 
on three datasets (BH76, RC21, and RSE43) 
self-consistently, with the HF density, and DC(HF).
Root-mean-squared-displacement of absolute errors
(RMSD) values are noted in the parenthesis.
$\tilde{S}_{avg.}$ is the averaged density sensitivity (kcal/mol) for the given dataset.
The deviation of HF's $\langle S^2 \rangle$ by more than 10\% from the ideal $\langle S^2 \rangle$ value is taken as a criterion for spin-contamination \cite{ADLM20}.
The D4 parameters uses the same parameter as PBE-D4 of Ref.~\cite{CEHN19}.
}
\label{tbl:spinc_all}
\end{table*}

\begin{figure*}[htb]
\centering
\includegraphics[width=2\columnwidth]{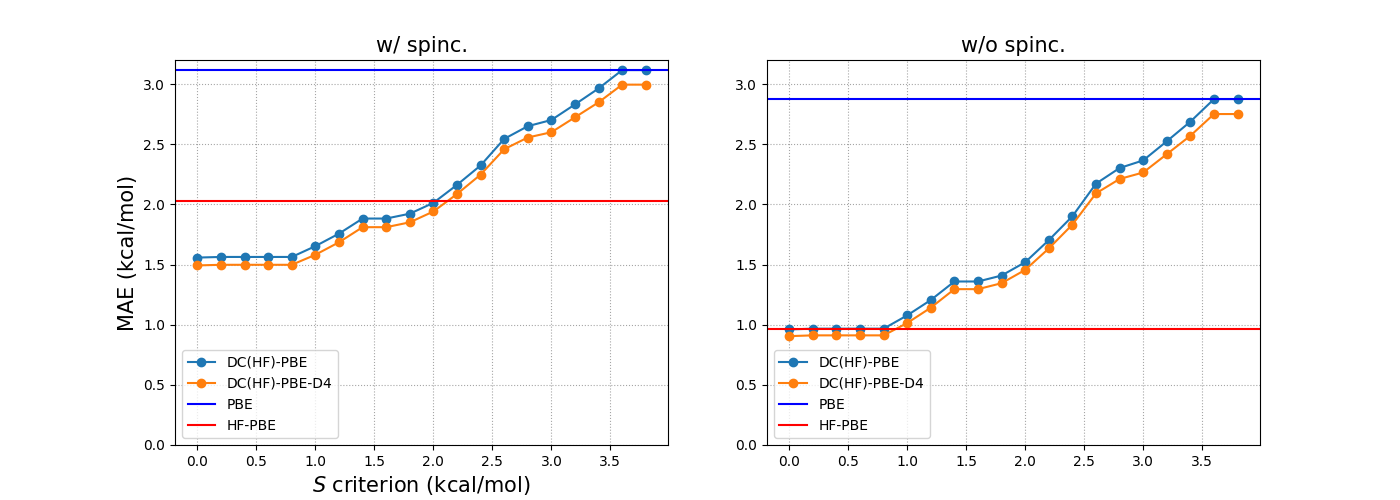}
\captionof{figure}[RSE43 MAE w.r.t. S criterion]{
MAE of RSE43 dataset with respect to the $\tilde S$ value cut-off criterion
for DC(HF)-PBE.
DC(HF)-PBE is SC-PBE for spin-contaminated cases and
below the $\tilde S$ cut-off criterion value (the x-axis of the plots) and HF-PBE otherwise.
The l.h.s.  panel is the MAE of all cases in RSE43 and 
the r.h.s.  panel is the MAE of non-spin-contaminated cases.
}
\label{fgr:rse43}
\end{figure*}

\begin{figure}[htb]
\centering
\includegraphics[width=1\columnwidth]{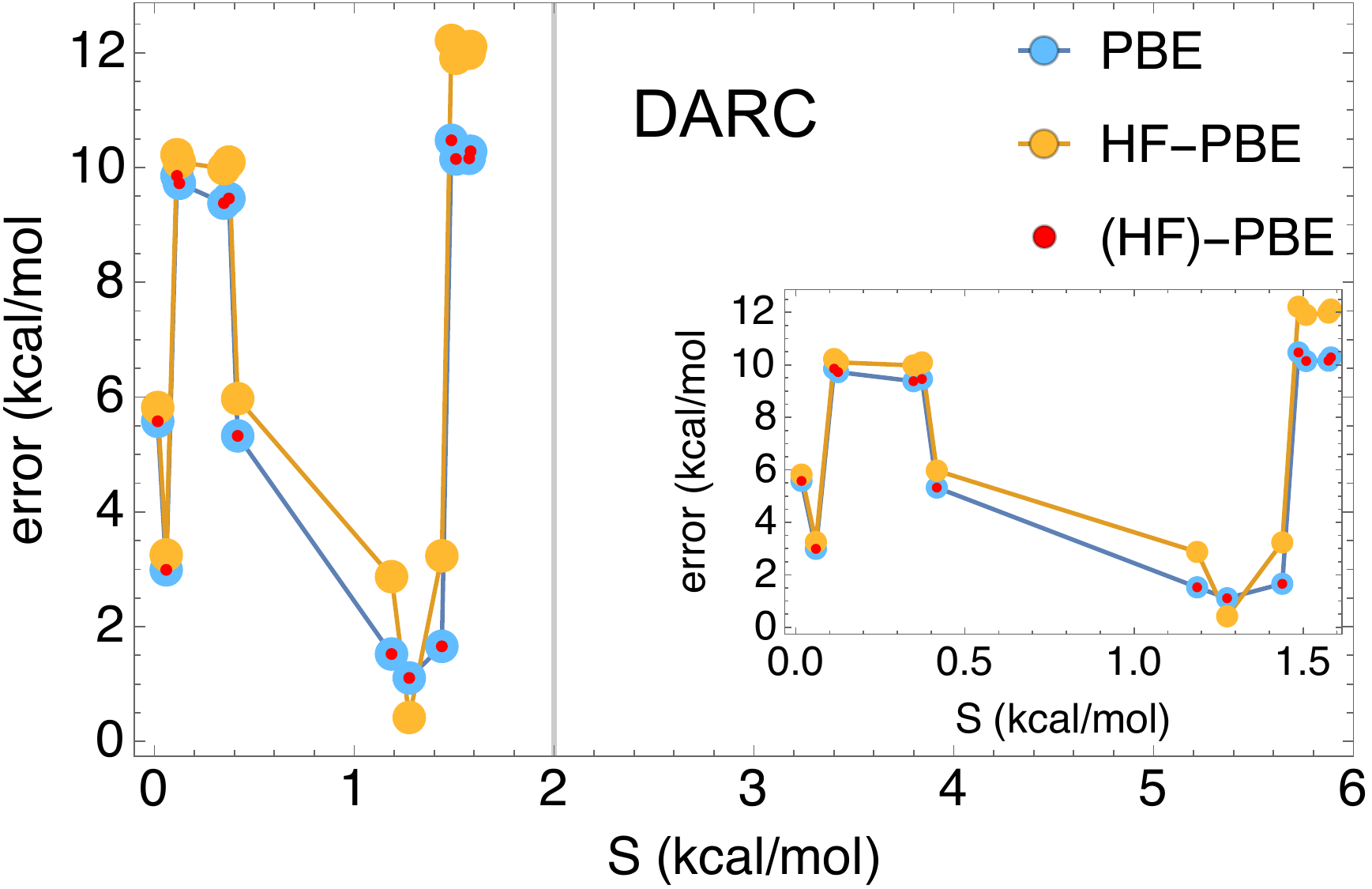}
\captionof{figure}[PBE and HF-PBE results for DARC]{
Same as Fig.~\ref{fgr:s},  but for the 
DARC dataset.}
\label{fgr:sd}
\end{figure}

\begin{figure}[htb]
\centering
\includegraphics[width=1\columnwidth]{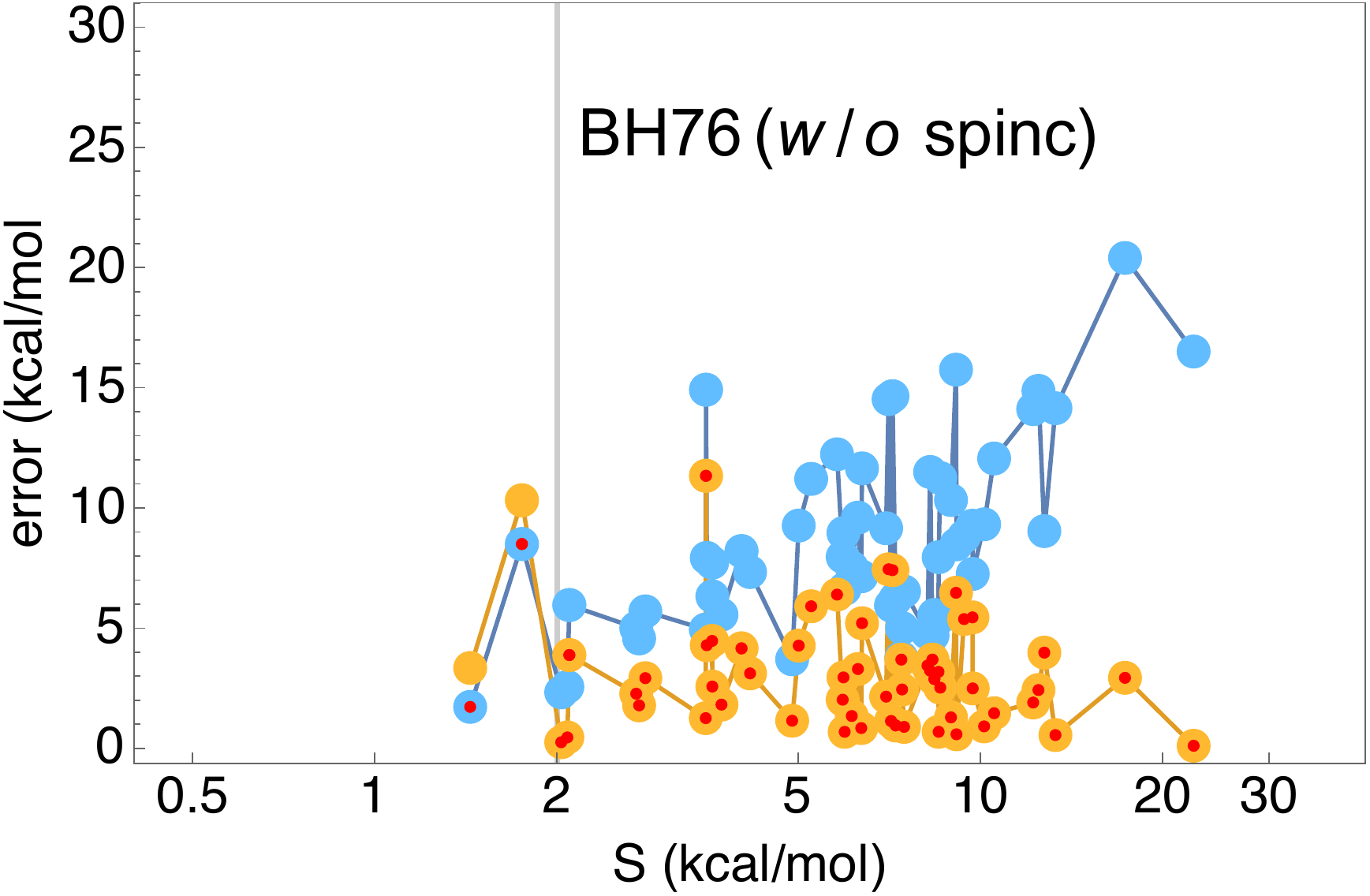}
\captionof{figure}[PBE and HF-PBE results for BH76 with spin-contaminated cases excluded]{
Same as Fig.~\ref{fgr:s},  but with spin-contaminated cases excluded.}
\label{fgr:sno}
\end{figure}

\begin{table*}[htb]
\centering
\resizebox{0.65\textwidth}{!}{
\begin{tabular}{l|cc|cc|ccccc}
& \multicolumn{4}{c}{Electron Affinity (kcal/mol)}                                                         & \multicolumn{3}{c}{$\epsilon_H$ (kcal/mol)} \\ \hline
\multicolumn{1}{l|}{}                           & \multicolumn{2}{c|}{PBE}                         & \multicolumn{2}{c|}{B3LYP}   & \multicolumn{4}{l}{}   \\
\multicolumn{1}{l|}{name}                           & \multicolumn{1}{c}{SC} & \multicolumn{1}{c|}{DC} & SC & \multicolumn{1}{c|}{DC} & PBE   & B3LYP   & HF & exact  \\ \hline
CCH	& 7.5	& 5.5	& 1.5	& -0.2	& 44.0	& 28.4	& -49.3	& -27.9 \\
CH$_2$	& 3.6	& 1.8	& 1.8	& 0.8	& 44.5	& 29.5	& -33.5	& -13.4 \\
CH$_3$	& 1.6	& -0.2	& -1.8	& -2.8	& 48.7	& 34.0	& -22.1	& -1.2 \\
NH	& 4.7	& 1.4	& 1.2	& -1.2	& 64.3	& 46.7	& 1.0	& -8.3 \\
NH$_2$	& 2.6	& 0.7	& -1.5	& -2.6	& 57.0	& 38.2	& -31.3	& -16.8 \\
OH	& 3.6	& 0.8	& -1.2	& -3.0	& 49.1	& 23.7	& -68.6	& -41.7 \\
SiH	& 2.8	& 2.2	& -1.9	& -2.9	& 28.7	& 18.9	& -35.7	& -29.3 \\
SiH$_2$	& 4.0	& 3.2	& -0.2	& -1.3	& 32.1	& 22.4	& -30.4	& -25.1 \\
SiH$_3$	& -0.2	& -0.1	& -1.9	& -1.6	& 20.4	& 6.3	& -44.2	& -31.4 \\
PH	& 1.0	& 0.6	& -0.1	& -0.3	& 38.0	& 24.4	& -21.3	& -23.5 \\
PH$_2$	& -0.2	& 0.0	& -2.0	& -1.8	& 35.1	& 21.4	& -28.7	& -28.8 \\
HS	& -0.6	& -0.7	& -2.5	& -2.4	& 22.0	& 4.3	& -59.7	& -54.2 \\
O$_2$	& -0.2	& -2.3	& 0.2	& -1.0	& 77.7	& 48.1	& -57.0	& -9.5 \\
NO	& 6.2	& 1.4	& 5.5	& 2.6	& 75.9	& 50.7	& -60.2	& 0.2 \\
CN	& -2.8	& 17.3	& 1.6	& 19.1	& 0.0	& -27.1	& -120.9	& -89.5 \\
PO	& 3.8	& 2.6	& 1.9	& 0.8	& 39.2	& 23.5	& -48.7	& -24.9 \\
S$_2$	& -2.0	& -1.5	& -1.5	& -1.1	& 28.5	& 10.4	& -53.0	& -38.0 \\
Cl$_2$	& 5.4	& 5.2	& 8.4	& 7.9	& -22.5	& -46.0	& -106.8	& -54.7 \\
\hline
MAE  & 2.9 (2.1)	& 2.6 (3.9)	& 2.0 (1.9)	& 3.0 (4.2)
                         &                                         &                                         \\
MAE*  & 2.9 (2.2)	& 1.8 (1.6)	& 2.1 (2.0)	& 2.0 (1.7)
                         &                                         &                                         \\
\hline \hline
\end{tabular}
}
\caption{
Same as the Table~\ref{tbl:g21ea} but with aug-cc-pvqz basis set.
Electron affinity errors for the G21EA dataset
relative to GMTKN55 reference.
HOMO energies (kcal/mol) for anions ($\epsilon_H$).
The exact value is the -EA of the reference energy.
MAE is the mean absolute error of all electrons affinities
while MAE* is the MAE without CN.
RMSD values are given in the paraenthesis.
CN is omitted due to large spin-contamination. 
See text for details.
}
\label{tbl:g21ea_sup}
\end{table*}


\begin{figure*}[htb]
\centering
\includegraphics[width=2.0\columnwidth]{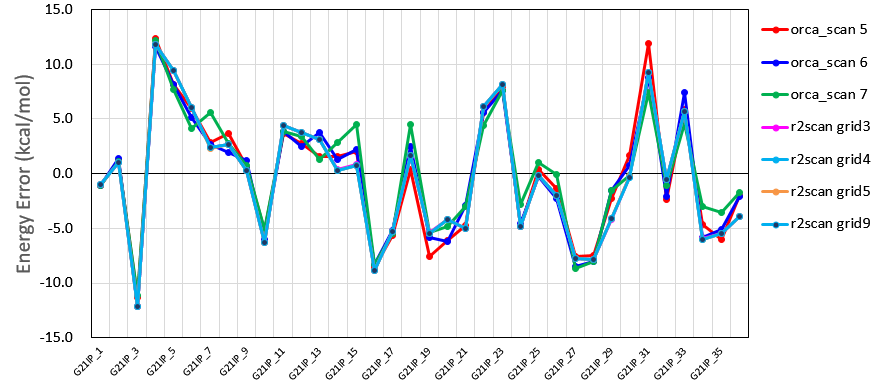}
\captionof{figure}[G21IP grid test2]{
G21IP reaction energy errors for SCAN and r2SCAN
with various grid levels.
SCAN shows a grid convergency issue (calculated by ORCA pacakge).
Note that all r2SCAN results are very similar (almost no changes).
}
\label{fgr:g21ip_grid_2}
\end{figure*}


\end{document}